\documentclass[10pt,nofootinbib,twocolumn,superscriptaddress,amsmath,amssymb,aps,pra]{revtex4-2}
\usepackage[table]{xcolor}
\usepackage{graphicx}
\usepackage{dcolumn}
\usepackage{bm}
\usepackage{xcolor}
\usepackage{mathtools}
\usepackage{multirow}
\usepackage{array}
\usepackage[normalem]{ulem}
\usepackage{enumitem}

\usepackage{orcidlink}

\usepackage{braket}
\usepackage{dsfont}

\renewcommand{\ket}[1]{\left|{#1}\right\rangle}

\newcommand{\ketbrad}[1]{\left|{#1}\rangle\!\langle{#1}\right|}

\usepackage{hyperref}

\newcommand{\mz}[1]{\textcolor{blue}{[{\tt MZ}: #1]}}

\newcommand{\GG}[1]{{\color{teal} [{\tt gg}: #1]}}

\definecolor{darkgreen}{RGB}{0,100,0}
\newcommand{\je}[1]{\textcolor{darkgreen}{[{\tt JE}: #1]}}

\begin{document}

	\title{Neural Network Learning of One-Bit Protocols for Qubit Measurement Simulation}
\author{Josep Escrig \orcidlink{0000-0002-0918-8148}}
\affiliation{Fundació i2CAT, internet i innovació digital a Catalunya, 08034 Barcelona, Spain}
\affiliation{Universitat Politècnica de Catalunya, 08005 Barcelona, Spain}
\author{Mani Zartab \orcidlink{0000-0003-2437-0906}}
\affiliation{Física Teòrica: Informació i Fenòmens Quàntics, Departament de Física,
	Universitat Autònoma de Barcelona, 08193 Bellaterra (Barcelona), Spain}
\author{Giulio Gasbarri \orcidlink{0000-0001-9135-1719} }
\affiliation{Naturwissenschaftlich-Technische Fakult{\"a}t, Universit{\"a}t Siegen, Siegen 57068, Germany}
\author{Estel Ferrer \orcidlink{0000-0001-9999-2844}}
\affiliation{Fundació i2CAT, internet i innovació digital a Catalunya, 08034 Barcelona, Spain}
\affiliation{Universitat Politècnica de Catalunya, 08005 Barcelona, Spain}
\author{Ramon Muñoz-Tapia \orcidlink{0000-0002-3048-9236}}
\affiliation{Física Teòrica: Informació i Fenòmens Quàntics, Departament de Física,
	Universitat Autònoma de Barcelona, 08193 Bellaterra (Barcelona), Spain}
\author{Gael Sentís \orcidlink{0000-0002-4982-6570}}
\affiliation{Física Teòrica: Informació i Fenòmens Quàntics, Departament de Física,
	Universitat Autònoma de Barcelona, 08193 Bellaterra (Barcelona), Spain}
	
	\date{\today}

\newpage
\begin{abstract}
Communication complexity provides a natural framework for quantifying the classical resources required to reproduce quantum statistics. In the qubit prepare-and-measure scenario, two classical bits have been shown to be  necessary and sufficient to simulate arbitrary qubit states and arbitrary quantum measurements exactly. However, this 
result does not exclude the possibility that restricted families of measurements may admit accurate 1-bit classical approximations. 
We use a neural network procedure to demonstrate that a single bit can achieve high average accuracy for specific measurement families. A performance analysis of our  neural network reveals that symmetric measurements with uniformly weighted elements, such as those forming regular polyhedra, are particularly amenable to this restricted communication. By analyzing the patterns learned by the neural network, we derive an analytical protocol that is 
extremely accurate for finite informationally complete symmetric configurations and becomes exact in the limit of a continuous isotropic measurement.

\end{abstract}
	
\maketitle

\section{Introduction}

Understanding how much classical communication is required to reproduce quantum statistics provides an operational way to compare quantum and classical information processing~\cite{brassard2003quantum,buhrman2010nonlocality}.
This perspective is especially relevant because quantum states serve as information carriers in communication protocols, cryptography, and dimension-limited information-processing tasks. In this setting, simulation costs provide a concrete benchmark for the expressive power of quantum communication: they indicate when a low-dimensional quantum system can encode correlations that would require substantially larger classical messages. 
The prepare-and-measure(PM) framework has been used to propose dimension witnesses~\cite{gallego2010device}, the certification of quantum devices~\cite{degois2021general},
the analysis of communication advantages~\cite{pawlowski2011qkd}, and the design and verification of quantum networks~\cite{bowles2015networks}, among other applications.

Several advances in this direction have shown that certain quantum statistics admit surprisingly efficient classical simulations. In Ref.~\cite{cerf2000classical}, the authors introduced a coding scheme for the classical teleportation of a qubit, requiring on average only 2.19 bits of classical information to reproduce the statistics of projective measurements. Toner and Bacon~\cite{toner2003communication} subsequently showed that shared classical randomness together with a single bit of communication suffices to reproduce the correlations of local projective measurements on a maximally entangled pair. In the PM scenario, they further showed that two classical bits are sufficient to simulate the statistics of arbitrary projective measurements on a qubit.
Degorre \emph{et al.}~\cite{degorre2005simulating} later observed that the communication step in the above protocol can be replaced by a source of biased shared randomness depending on the qubit state, thereby identifying a key ingredient underlying classical simulations of quantum statistics. Building on these ideas, Renner \emph{et al.}~\cite{renner2023classical} extended the simulation of qubit statistics from projective to generalized measurements, proving that any qubit PM scenario can be simulated exactly with a worst-case communication cost of two bits. Extensions to higher-dimensional systems have also been investigated~\cite{montina2011approximate,rudolph2006ontological,zartab2026prepare,schlosser2026bounding}, including approximate simulation protocols and analyses of average communication cost. Ultimately, the existence of an exact classical simulation protocol for PM scenarios in dimension $d \geq 3$ using a finite amount of classical communication remains open.

The 2-bit worst-case simulation of arbitrary qubit PM statistics, however, does not preclude more economical descriptions in less demanding regimes. One natural relaxation is to move from worst-case to average communication cost. Indeed, recent work has shown that, by encoding the messages sent by Alice, qubit PM statistics can be simulated with an average communication cost of 1.89 bits~\cite{schlosser2026bounding}. This suggests that the apparent 2-bit barrier may be a consequence of demanding a uniform worst-case guarantee over all state preparations and measurements, and that more efficient simulations may be possible once the relevant operational figure of merit is adapted.

A second route toward lower communication cost is to restrict the family of quantum states and/or measurements to be simulated. This is particularly natural in PM scenarios, where practical communication tasks often involve structured ensembles of preparations and measurements rather than the full set of qubit statistics. From this perspective, the central question becomes which physically or operationally meaningful subsets of qubit statistics already admit an even more efficient classical simulability. Identifying such subsets would help clarify which features of quantum PM experiments are responsible for their classical communication cost.

Further motivation comes from recent numerical evidence in the Bell scenario. In Ref.~\cite{sidajaya2023neural}, neural network methods were used to search for 1-bit classical protocols simulating the statistics of local projective measurements on bipartite entangled two-qubit states. Strikingly, although one bit of communication is known to be insufficient in full generality, the authors found that 1-bit protocols approximate the target quantum correlations extremely well across the tested instances, and no explicit counterexample was identified by their search. This suggests that one bit of communication can already capture a surprisingly large portion of quantum statistics, extending the intuition provided by the Toner--Bacon protocol beyond the maximally entangled singlet case~\cite{toner2003communication}.

These observations motivate the search for 1-bit simulation protocols in PM scenarios. In this work, we focus on restricted families of measurements and ask whether their qubit statistics can be reproduced, or accurately approximated, using only a single bit of classical communication. To this end, we adopt a neural network(NN)-~procedure, in the spirit of Ref.~\cite{sidajaya2023neural}, using the network as a tool to discover candidate protocols rather than imposing a fixed analytical ansatz from the outset. When successful, such searches provide evidence for hidden structure in the corresponding family of measurements and point toward compact classical descriptions of their quantum statistics. Thus, our goal is to map regimes of 1-bit-simulability in qubit PM scenarios and to use protocols, discovered by NN-procedure, as evidence for new analytically tractable classical simulations.

Following this approach, we first train a neural network on randomly generated PM instances
and use it to identify structured families of measurements that are strong candidates for 1-bit simulability. The families singled out by the optimization are highly-symmetric POVMs whose elements are distributed close to uniformly on the Bloch sphere. By inspecting the behavior learned by the network, we then extract an explicit analytical protocol, thereby moving from a black-box numerical search to a concrete classical simulation strategy.

Our analysis shows that the extracted 1-bit protocol, while it is not exact for arbitrary finite POVMs, yields highly accurate approximations to the quantum probabilities for informationally-complete symmetric POVMs, and the approximation improves as the number of outcomes increases. In the continuous limit, corresponding to the covariant qubit POVM, the protocol becomes exact, placing this measurement within the class of exactly 1-bit-simulable PM resources. These results provide evidence that symmetry can substantially reduce the communication needed to simulate quantum statistics, and they illustrate how neural network searches can help uncover analytical structure in PM simulation problems.

This paper is structured as follows. In Section~\ref{definitions and background} we introduce the theoretical background and definitions used throughout this paper. In Section~\ref{NN-procedure} we describe the NN-procedure used to learn quantum statistics under the restrictions of classical shared randomness and one bit of classical communication, and in Section~\ref{sec:num-perf} we present the numerical results. In Section~\ref{sec:classical_protocol} we infer the classical 1-bit protocol and study its properties. We end with the conclusions and include the technical details in the Appendices.


\section{Background}
\label{definitions and background}

A general $m$-outcome quantum measurement is described by a POVM, \emph{i.e.}, a set of positive semidefinite operators $\mathcal{M}:=\{M_i\}_{i=1}^m$ satisfying the completeness condition $\sum_{i=1}^m M_i = \openone$. 
The probability of obtaining outcome $i$ when applying $\mathcal{M}$ on a generic state $\rho$ is given by the Born rule $P_Q(i| \rho,\mathcal{M}) = \mathrm{Tr}(M_i\rho)$. For qubits, each POVM element can be written in the Bloch representation as 
$M_i = p_i\bigl(\openone+\vec y_i\cdot\vec\sigma\bigr)$,
where $p_i\ge 0$, $\sum_i p_i=1$, and $|\vec y_i|\le 1$. Similarly, we can express a qubit state as $\rho= (\openone + \vec{r}\cdot \vec{\sigma})/2$.

The probability $P_Q(i | \rho,\mathcal{M})$ then takes the form
\begin{align}
    P_Q(i | \rho,\mathcal{M}) = p_i\bigl(1+\vec y_i\cdot\vec r\bigr).
    \label{eq:born_bloch}
\end{align}

\begin{figure}[t]
    \centering
    \includegraphics[width=8cm]{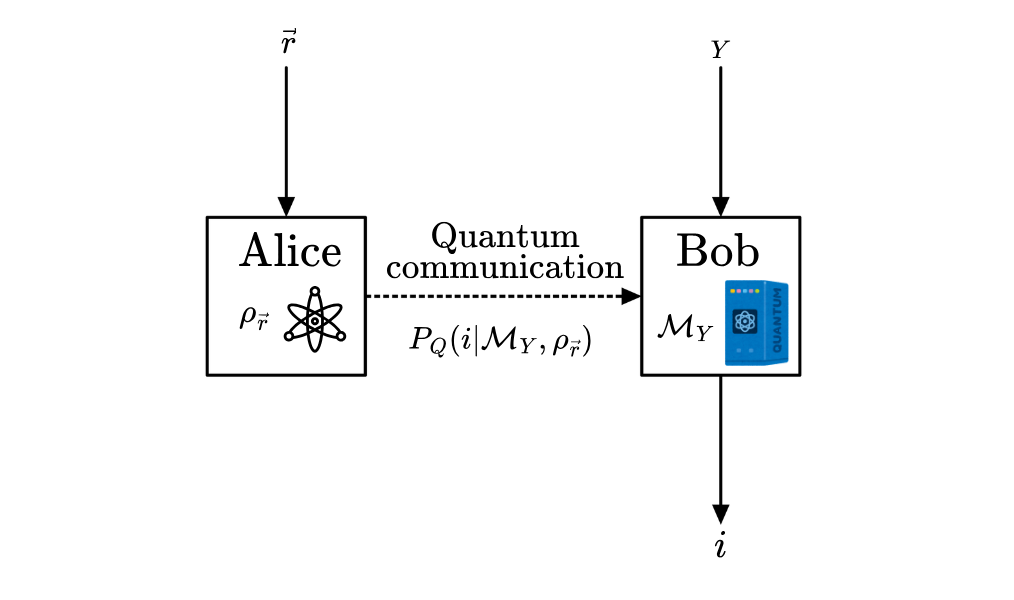}
    \caption{Quantum PM scenario. Alice prepares the qubit state $\rho_{\vec r}$ and sends it to Bob, who performs the POVM $\mathcal{M}_{Y}$ 
    and reports the outcome $i$.}
    \label{qpm-scenario}
\end{figure}

Let us now introduce the quantum PM scenario. In this setup, Alice prepares a qubit state $\rho_{\vec r}$, represented by its Bloch
vector $\vec r$, and sends it to Bob through a quantum communication channel.
Bob then performs a POVM $\mathcal{M}_{Y}$ specified by 
$Y:=(p_i,\vec y_i)_{i=1}^m$, 
and reports the outcome $i$ with probability given by (\ref{eq:born_bloch}). The scenario is illustrated in Fig.~\ref{qpm-scenario}. 

A classical simulation replaces the transmitted qubit by a classical message, assisted by shared randomness.
Alice receives the state description $\vec{r}$, Bob receives the measurement description $Y$, and they share a random variable independent of both. Alice sends a classical messace $c$ to Bob. Bob, uses $c$, $Y$, and the share randomness to generate an outcome $i$. 
The protocol exactly simulates the quantum PM scehnario if, for all states and measurements, the resulting classical probability $P_{C}(i|\vec{r},Y)$ coincides with the quantum one \textit{i.e},
\begin{align}
P_{C}(i|\vec{r},Y) = P_{Q}(i|\rho_{\vec{r}},\mathcal{M}_Y) \quad \forall i,\vec{r}, Y.
\end{align}

If the equality holds only up to a chosen error metric, evaluated on a spceific class of states and measurements, we will refer the protocol as an approximate simulation.

\begin{figure}[t]
    \centering
    \includegraphics[width=8cm]{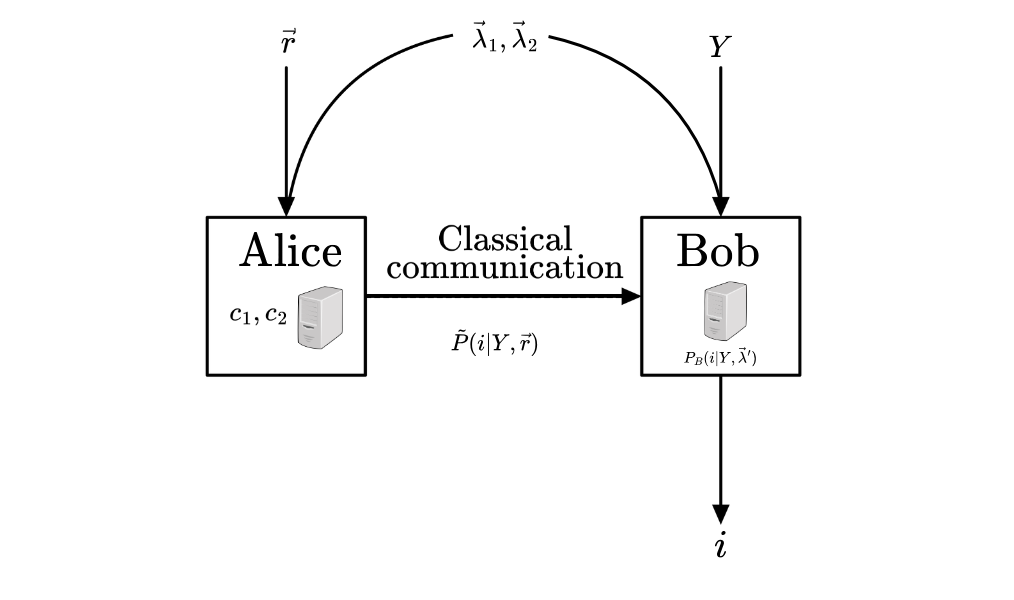}
    \caption{2-bit classical simulation protocol of Ref.~\cite{renner2023classical} for the PM scenario. Alice and Bob share $\vec\lambda_1,\vec\lambda_2\in\mathbb S^2$, Alice sends the bits $c_1,c_2$, and Bob outputs $i$ according to the protocol.}
    \label{simulation}
\end{figure}

Renner \emph{et al.}~\cite{renner2023classical} showed that the qubit PM scenario admits an exact classical simulation with two bits of communication, and that two bits are necessary in the worst case. For comparison with the results of the following sections, we briefly recall their protocol. 
Without loss of generality, it is enough to consider pure states, $|\vec{r}|=1$, since mixed states can be decomposed as a convex combination of pure states, and the corresponding classical randomness can be incorporated into the shared randomness of the protocol. Similarly, for measurements, it is enough to consider POVM elements proportional to rank-one projectors, (\emph{i.e.}, $\left|\vec{y}_i\right| = 1,\, \forall i$), since general qubit POVMs can be obtained from rank-one refinements by classical post-processing. 
The protocol is illustrated in Fig.~\ref{simulation} and proceed as follows:

\begin{enumerate}
    \item Alice is given the Bloch vector $\vec{r}$, and Bob is given the POVM  specified by $Y$. In addition, they share two random unit vectors   $\vec\lambda_1,\vec\lambda_2\in\mathbb S^2$.

    \item Alice computes the classical bits
    \begin{align}
        c_1 = H(\vec\lambda_1\cdot\vec r), \qquad
        c_2 = H(\vec\lambda_2\cdot\vec r),
    \end{align}
    and sends them to Bob, where $H(x)$ is the Heaviside step function.

    \item Bob defines the effective hidden variables
    \begin{align}
        \vec\lambda_i^\prime = (2c_i-1) \vec\lambda_i,
    \end{align}
    that is, he flips $\vec\lambda_i$ whenever $c_i=0$.

    \item Bob selects $\vec y_i$ with probability $p_i$. He then sets
    $\vec\lambda^\prime =\vec\lambda_1^\prime$ if
    $|\vec\lambda_1^\prime\cdot\vec y_i|\geq |\vec\lambda_2^\prime\cdot\vec y_i|$,
    and $\vec\lambda^\prime=\vec\lambda_2$ otherwise.

    \item Bob reports the outcome $i$ with probability
    \begin{align}
        P_B(i| \vec\lambda^\prime,Y )=
        \frac{p_i(\vec y_i\cdot\vec\lambda^\prime)\,H(\vec y_i\cdot\vec\lambda^\prime)}
        {\sum_{j=1}^m p_j(\vec y_j\cdot\vec\lambda^\prime)\,H(\vec y_j\cdot\vec\lambda^\prime)}.
        \label{eq:renner_bob_rule}
    \end{align}
\end{enumerate}

As shown in Ref.~\cite{renner2023classical}, the statistics generated by this protocol can be written as
\begin{align}
    P_C(i| \vec r,Y)
    =
    \int_{\mathbb S^2} d\vec\lambda\;
    \rho(\vec\lambda| \vec r,Y)\,
    P_B(i| \vec\lambda,Y),
    \label{rtq}
\end{align}
where $d\vec{\lambda}=d\Omega/(4\pi)$ denotes the normalized uniform measure on the sphere. The effective distribution $\rho(\vec{\lambda}|\vec{r},Y)$, which encodes the statistical effect of the flipping and comparison steps, is given by:
\begin{equation}
  \rho(\vec\lambda| \vec r,Y)=8 H(\vec{r}\cdot\vec{\lambda})
  \sum_{j=1}^m 
  p_j(\vec y_j\cdot\vec\lambda)\,H(\vec y_j\cdot\vec\lambda).
  \label{eq:prob-final-renner}
\end{equation}

\section{Neural Network learning procedure}

\label{NN-procedure}
Another important result of Ref.~\cite{renner2023classical} is that no classical protocol using fewer than two classical bits is able to reproduce the statistics of all possible quantum input setups in this setting. 

However, this does not exclude the possibility that restricted families of measurements may admit simpler classical simulations. \emph{A priori}, it remains unclear which structural properties of a POVM may enable such a reduction, and which protocols could possibly achieve it. Our aim is 

to use a NN-procedure, as a data-driven probe for identifying such regimes.

For a given POVM $\mathcal{M}_{Y}$, Alice and Bob share a random unit vector $\vec\lambda\in\mathbb S^2$, sampled uniformly from the Bloch sphere. Alice receives the input state $\rho_{\vec r}$ and sends to Bob the single bit $ c = H(\vec\lambda\cdot\vec r)$.
The neural network takes as input the pair $(c,\vec\lambda)$ and outputs a probability distribution over the outcomes of the POVM, $P_{\rm{NN}}(i|c,\vec{\lambda},Y)$.
This output plays the role of Bob's response function. The dependence on $Y$ is implicit in the trained network: for each POVM $\mathcal {M}_{Y}$ considered, a separate network is trained.
The statistics generated by the NN-procedure are obtained by averaging the response function over the shared randomness. For a fixed input state $\rho_{\vec r}$, the averaged prediction is
\begin{align}
P_{\rm{NN}}(i|\vec{r},Y)=\int_{\mathbb S^2} d\vec{\lambda}\: P_{\rm{NN}}\bigl(i|H(\vec{\lambda}\cdot\vec r),\vec{\lambda},Y\bigr).
\label{eq:nn_ideal_average}
\end{align}
In practice, this integral is estimated by Monte Carlo sampling.

The training set consists of $M$ Haar-random pure input states $\{\vec{r}^{(j)}\}_{j=1}^M$. For each state $\vec{r}^{(j)}$, we sample $N$ independent shared random vectors $\{\vec{\lambda}^{(j,k)}\}_{k=1}^N$ and define the corresponding bits $
c^{(j,k)}=H\bigl(\vec\lambda^{(j,k)}\cdot\vec r^{(j)}\bigr)$.
For each state $j$ and outcome $i$, the Monte Carlo average of the neural network response is
\begin{align}
\bar p_{i}^{(j)}
=
\frac{1}{N}\sum_{k=1}^{N} P_{\rm{NN}}\bigl(i|c^{(j,k)},\vec{\lambda}^{(j,k)},Y\bigr).
\label{eq:nn_mc_average}
\end{align}
The target probability is the Born probability
\begin{align}
P_Q(i|\rho_{\vec{r}^{(j)}},\mathcal{M}_{Y})
=
p_i\bigl(1+\vec y_i\cdot\vec r^{(j)}\bigr).
\label{eq:nn_target}
\end{align}

The network is trained by minimizing the mean squared error(MSE) between the averaged neural network probabilities and the corresponding Born probabilities,
\begin{align}
\mathcal L_{\rm{MSE}}
=
\frac{1}{M}\sum_{j=1}^{M}\sum_{i=1}^{m}
\left[\bar p_{i}^{(j)}-P_Q(i|\rho_{\vec{r}^{(j)}},\mathcal{M}_Y)\right]^2.
\label{eq:nn_mse_loss}
\end{align}
We use MSE as the training loss because it is smooth and penalizes large deviations more strongly, which is convenient for numerical optimization.

To evaluate the accuracy of the trained network, we use the mean absolute error(MAE),
\begin{equation}
\mathrm{MAE}
=
\frac{1}{M}\sum_{j=1}^{M}\sum_{i=1}^{m}
\left|
\bar p^{(j)}_i - P_Q(i| \rho_{\vec r^{(j)}}, \mathcal{M}_{Y})
\right|.
\label{eq}
\end{equation}
Unlike the MSE, the MAE directly quantifies the average absolute discrepancy between the predicted and target probabilities, providing a more standard figure of merit for comparing the accuracy of different protocols.

Two important remarks are in order. First, for each POVM whose statistics we wish to simulate, the neural network must be retrained. Nevertheless, as we show below, the analysis of the resulting network responses allows us to infer a POVM-independent 1-bit protocol. Second, because the MAE is evaluated over randomly sampled states and shared random vectors, it quantifies an \emph{average} simulation accuracy. The training methodology and network architecture are described in more detail in Appendix~\ref{app:training}.

\section{Numerical performance}\label{sec:num-perf}

We now analyze the numerical performance of the neural network, introduced in the previous section, with a focus on how it varies depending on the characteristics of the POVMs.

Throughout the numerical experiments, in this section, unless otherwise stated, we fix the parameters to $N=4000$ shared randomness samples and $M=2000$ input states during the training stage and $N=4000$, $M=1000$ during the test stage. We found these values to provide a good compromise between numerical accuracy and computational cost. Additional implementation details are given in Appendix~\ref{app:training}.

We first study the NN-procedure on randomly generated POVMs with different numbers of outcomes. Random POVMs can be sampled in several inequivalent ways, depending on the chosen measure over the space of effects~\cite{heinosaari2020randompovm}. In the present work, we use a constructive procedure based on random Bloch-sphere directions followed by a numerical search for nonnegative weights satisfying the POVM completeness condition.

\begin{figure}[h]
    \centering
    \includegraphics[width=\columnwidth]{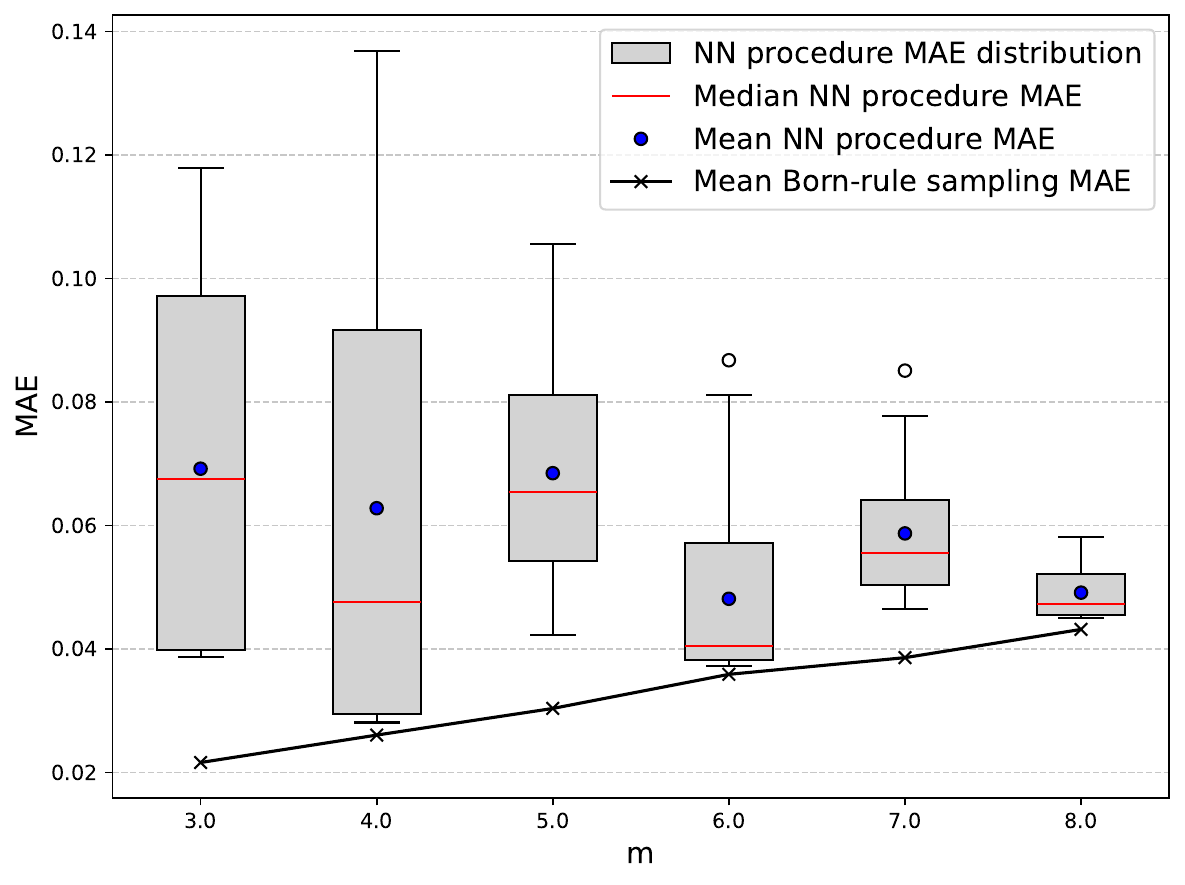}
    \caption{MAE of the probabilities calculated with the NN-procedure for POVMs with different numbers of elements $m$, compared with the mean MAE sampled directly from the Born rule.}
    \label{quartile}
\end{figure}

Fig.~\ref{quartile} summarizes the MAE obtained in $100$ experiments over $100$ different randomly generated POVMs with numbers of outcomes ranging from $m=3$ to $m=8$.

The figure reports the median, mean, and interquartile range of the MAE values obtained with the NN-procedure.
For comparison, we also show the MAE obtained by direct finite-sample estimation of the Born probabilities.
This Born rule curve should be interpreted as a sampling baseline: it quantifies the error due to finite sampling when the target probabilities are known.

The most immediate observation is that the MAE of the NN-procedure is systematically larger than the MAE obtained by direct Born rule sampling. Nevertheless, the results also show that some POVM instances are reproduced much more accurately than others. In particular, for certain values of $m$ the best cases approach the Born rule baseline rather closely. This suggests that the performance of the 1-bit protocol 
may depend on structural properties of the measurement.

\begin{figure}[htbp]
    \centering
    \includegraphics[width=\columnwidth]{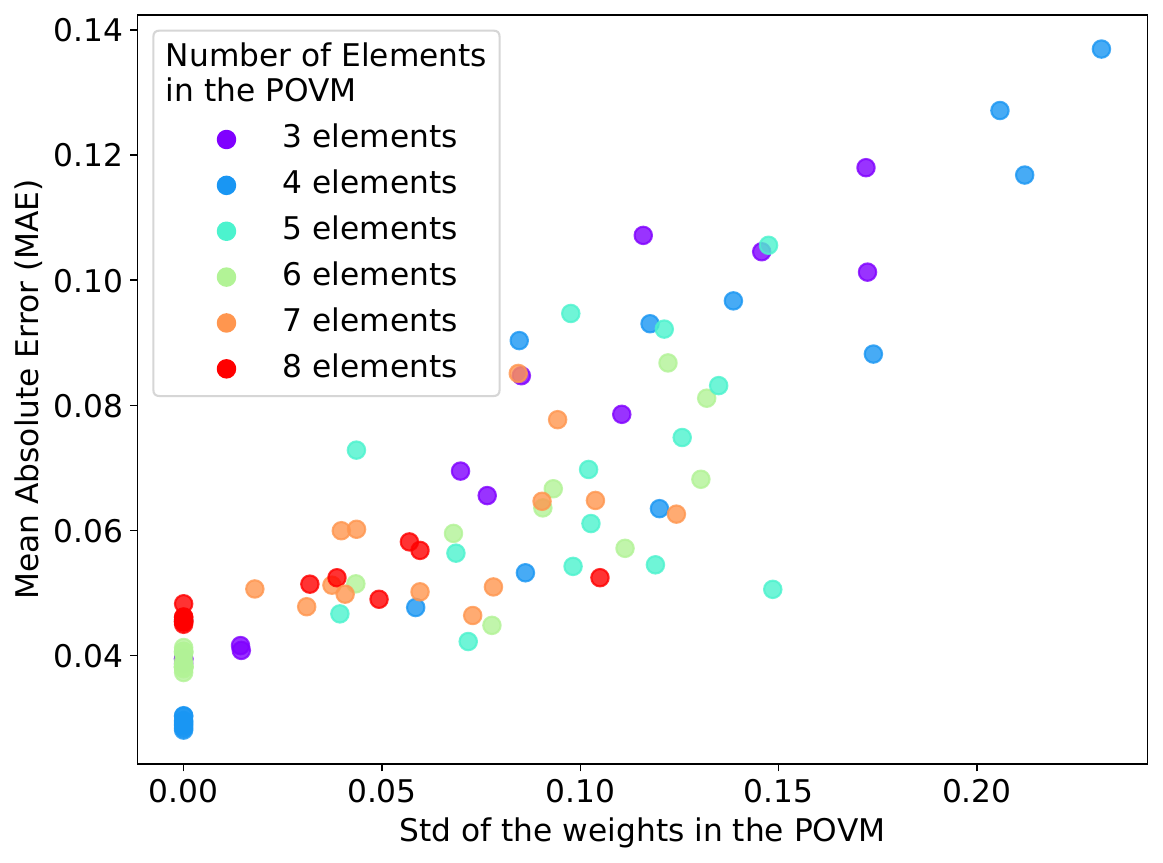}
    \caption{MAE of the probabilities obtained from the NN-procedure for different POVMs as a function of $\mathrm{Std}$ of the weights $\{p_i\}_{i=1}^{m}$ of the elements of each POVM. POVMs with different numbers of elements $m$ are represented with different colors.}
    \label{povmspovm}
\end{figure}

By direct inspecting the well-performing POVMs, we observe that the uniformity in the weight $\{p_i\}_{i=1}^m$ distribution across the POVM elements seems to play an important role. To investigate this hypothesis, we plot the MAE as a function of the standard deviation of the POVM weights. A smaller standard deviation corresponds to a more uniform distribution of the weights, with $\mathrm{Std}=0$ corresponding to the equal-weight case $p_i=1/m$ for all $i$.
Fig.~\ref{povmspovm} shows that the accuracy of the NN-procedure improves significantly as this standard deviation decreases, for any number of outcomes $m$. The most accurate cases are concentrated near the standard deviation $\mathrm{Std}=0$.

Motivated by this numerical trend, we focus from this point onwards on POVMs with equal weights, \textit{i.e.}, $p_i=1/m\,,\forall i$.
In the rank-one case, these measurements can be written as $\left\{ \frac{2}{m}\ketbrad{\psi_i}\right\}_{i=1}^{m},$ where each $\ket{\psi_{i}}$ is a normalized vector. The POVM normalization condition is therefore $\sum_{i=1}^m \ketbrad{\psi_i} = (m/2) \openone$, which implies that the vectors $\{\ket{\psi_i}\}_{i=1}^m$ form a unit-norm tight frame.
We refer to this class as \emph{equal-trace rank-one POVMs} or e-POVMs.
Accordingly, they have also been termed unit-norm tight-frame POVMs in the literature~\cite{eldar2002frames,casazza2013finite}.
Geometrically, under the Bloch-sphere representation, the pure states associated with an e-POVM correspond to $m$ points on the sphere whose centroid lies at the origin. Thus, the POVM normalization condition admits a simple interpretation as a balance condition on the corresponding Bloch-sphere configuration $\sum_{i=1}^{m} \vec{y}_{i}=\vec{0}$, with $|y_{i}|=1$, where the POVM elements take the form $M_{i}=\frac{1}{m}\left(\openone+\vec{y}_i\cdot\vec{\sigma}\right)$.

When needed, we refer to the informationally complete(IC) members of this class as eIC-POVMs.

Within the class of e-POVMs, it is natural to examine highly regular configurations~\cite{slomczynski2014hspovms} of Bloch vectors. 
We therefore compare the performance of the NN-procedure on regular polygons and regular polyhedra inscribed in the Bloch sphere. Both classes have equal weights and balanced Bloch vectors, but they differ in an important respect: regular polygons lie in a plane and are therefore not informationally complete, whereas regular polyhedral configurations span $\mathbb{R}^{3}$ and define eIC-POVMs.
The set of regular polygons and polyhedra inscribed in the Bloch sphere\footnote{These structures also belong to the class of "highly-symmetric POVMs" defined in Ref.~\cite{slomczynski2014hspovms}.}

\begin{figure}[t]
    \centering
    \includegraphics[width=0.49\columnwidth]{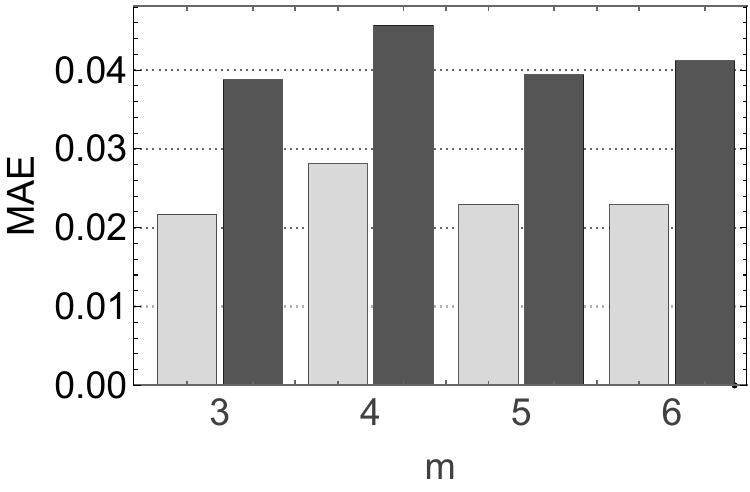}
    \hfill
    \includegraphics[width=0.49\columnwidth]{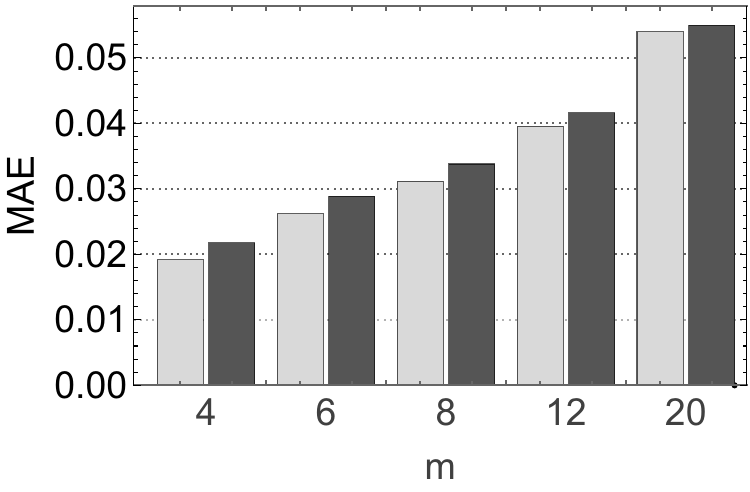}
    \caption{MAE of the NN-procedure (dark gray) and of the Born rule (light gray) for various e-POVMs with $m$ outcomes, taking 4000 samples (value of $N$) and averaging over 1000 different states (value of $M$). In the left plot, we have polygon configurations, while on the right plot we have regular polyhedra.}
    \label{fig:bars}
\end{figure}

Fig.~\ref{fig:bars}
compares the MAE of the NN-procedure with the MAE obtained by direct sampling from the Born rule for several such configurations. The comparison reveals a clear qualitative distinction. For polygonal configurations such as the triangle, square, pentagon, and hexagon, the MAE of the NN-procedure remains noticeably above the Born rule baseline. By contrast, for regular polyhedra configurations, the gap between the NN-procedure and the Born rule baseline is significantly smaller. This suggests that the symmetric three-dimensional configuration of measurement vectors, representing an informationally complete POVM plays an important role.

Overall, the numerical results support two main observations. First, 1-bit NN-procedure can reproduce the statistics of certain POVMs with high average accuracy, even though one bit cannot simulate the full qubit PM scenario exactly.
Second, this performance is not uniform across all measurements. It strongly depends on the distribution of the POVM weights. In particular, the protocol performs best for 

eIC-POVMs in highly symmetric configurations. 

\section{Inferred classical 1-bit protocol}
\label{sec:classical_protocol}
The numerical results of the previous section indicate that the NN-procedure performs particularly well for eIC-POVMs whose Bloch vectors form highly isotropic three-dimensional configurations. This motivates the search for a transparent analytical description of the behavior learned by the neural network.

We first inspect the neural-network response and extract a simple analytical rule. We then define the corresponding 1-bit protocol and study its numerical performance for finite eIC-POVMs. Finally, we show that the protocol, although not exact for finite eIC-POVMs in general, becomes exact for regular polyhedral configurations when the input state aligns with one of the measurement outcomes, and also, in the continuous isotropic limit.

\subsection{Extraction and definition of the 1-bit protocol} 
\label{subsec:extraction_definition_protocol}

To identify the structure learned by the NN-procedure, we inspect the neural-network output for a representative eIC-POVM.

We use the six-direction Cartesian configuration, namely the POVM whose Bloch vectors are $\pm \hat{x}$, $\pm \hat{y}$, and $\pm \hat{z}$. This is the octahedral configuration on the Bloch sphere and corresponds to the union of the three mutually unbiased qubit bases.

\begin{figure}
    \centering
    \includegraphics[width=1\linewidth]{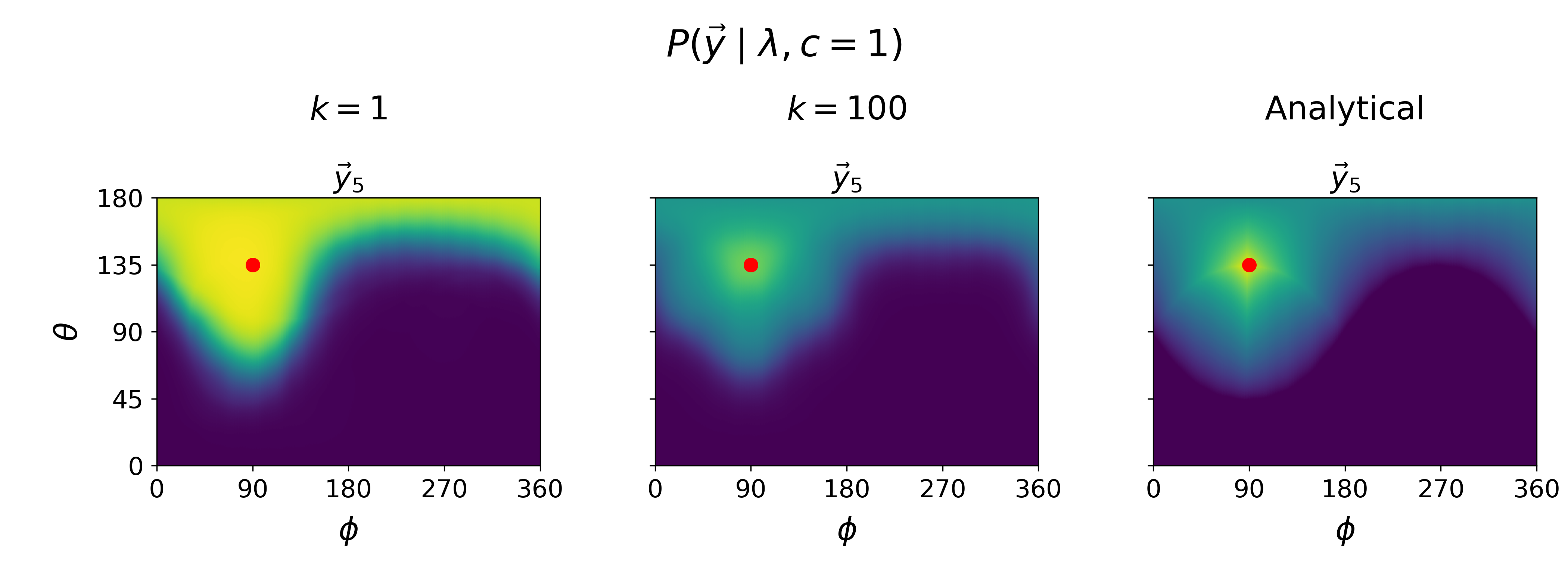}
    \caption{Conditional response probabilities for outcome $i=5$ of the six-direction Cartesian, or octahedral, eIC-POVM, conditioned on a transmitted bit $c=1$. The left panel shows the output of a single trained neural network ($k=1$), the center panel shows the output averaged over an ensemble of $k=100$ independently trained networks, and the right panel shows the response of the analytical 1-bit protocol in (\ref{eq:semi_analytical_protocol}). The probabilities are represented as a function of the shared random vector $\vec{\lambda}$, parametrized by the polar angles $(\phi,\theta)$. The red marker indicates the Bloch vector $\vec y_5$ associated with the selected outcome.}
    \label{fig:placeholder}\label{Pnn_map_3panels}
\end{figure}

Fig.~\ref{Pnn_map_3panels} shows the conditional response of the neural network for the outcome associated with $+\hat z$, and for the transmitted bit $c=1$, $P_{\mathrm{NN}} (i=5|c=1, \vec\lambda,Y)$, as a function of the shared vector $\vec\lambda$, parametrized by the polar angles $(\theta,\phi)$. The red dot indicates the polar direction of the POVM element $\vec{y}_5$. 
The left panel shows the response of a single trained neural network. 
Although the response is smooth and shows a clear geometric dependence on $\vec{\lambda}$, the fluctuations associated with a particular training run prevent sharp boundaries between the decision regions, making difficult to infer a simple analytical rule directly. 
To reduce these network-dependent fluctuations, we average the response probabilities over $100$ independently trained networks. The resulting ensemble-averaged response is shown in the central panel.
The corresponding response maps for all outcomes and for both values of the communicated bit are provided in Appendix~\ref{app:additional_plots}.

The ensample-averaged response reveals a simple pattern. For $c=1$, the probability associated with the POVM element $\vec{y}_i$ is approximately zero on the hemisphere opposite to $\vec{y}_i$, and increases approximately linearly with the scalar product $\vec{y}_i\cdot\vec{\lambda}$ on the hemisphere centered on $\vec{y}_i$. This suggests the proportionality rule
\begin{equation}
P_{\mathrm{NN}}(i|c=1, \vec{\lambda},Y)
\propto
(\vec{y}_i\cdot\vec{\lambda})\,H(\vec{y}_i\cdot\vec{\lambda}),
\label{eq:nn_rule_c1}
\end{equation}

For $c=0$, the same pattern is observed after the flip of the hidden variable $\lambda$, \emph{i.e.}, $\vec{\lambda}\mapsto -\vec{\lambda}$, leading to
\begin{equation}
P_{\mathrm{NN}}(i| c=0,\vec{\lambda},Y)
\propto
(-\vec{y}_i\cdot\vec{\lambda})\,H(-\vec{y}_i\cdot\vec{\lambda}).
\label{eq:nn_rule_c0}
\end{equation}

Equations~\eqref{eq:nn_rule_c1} and \eqref{eq:nn_rule_c0} suggest that the dependence on the communicated bit can be described by a sign flip of the shared vector $\lambda$. This motivates the following 1-bit protocol:

\begin{enumerate}
\item Alice receives the description of a qubit state $\rho_{\vec{r}}$, represented by its Bloch vector $\vec{r}$, and Bob receives the description of an e-POVM $\mathcal{M}_{Y}$ specified by $Y$ (Remember, $p_i=1/m$ $\forall i$).
In addition, they share a random unit vector $\vec{\lambda}\in \mathbb{S}^2$, uniformly distributed over the Bloch sphere.

\item Alice computes the bit
\begin{equation}
c = H(\vec{\lambda}\cdot\vec{r}),
\label{eq:sa_bit}
\end{equation}
and sends it to Bob.

\item Bob defines the effective hidden variable
\begin{equation}
\vec{\lambda}^{\,\prime}=(2c-1)\vec{\lambda}.
\label{eq:sa_flip}
\end{equation}
Thus, Bob flips the shared vector whenever $c=0$.

\item Bob reports the outcome $i$ with probability
\begin{equation}
P_{\mathrm{1-bit}}(i| \vec{\lambda}^{\,\prime},Y)
=
\frac{(\vec{y}_i\cdot\vec{\lambda}^{\,\prime})\,H(\vec{y}_i\cdot\vec{\lambda}^{\,\prime})}
{\sum_{j=1}^{m}(\vec{y}_j\cdot\vec{\lambda}^{\,\prime})\,H(\vec{y}_j\cdot\vec{\lambda}^{\,\prime})}.
\label{eq:semi_analytical_protocol}
\end{equation}
\end{enumerate}

The right panel of Fig.~\ref{Pnn_map_3panels} shows the response of (\ref{eq:semi_analytical_protocol}) for the octahedral outcome associated to $i=5$.
Its geometric structure closely reproduces the ensamble-averaged neural network response: the probability of outcome $i$ is maximal at the Bloch vector $\vec{y}_i$ and concentrated in the hemisphere defined by it and increase with the positive overlap $\vec{y}_i\cdot\vec{\lambda'}$.

\subsection{Numerical performance}
\label{subsec:numerical_performance_sa}

We tested the analytical protocol in (\ref{eq:semi_analytical_protocol}) on six e-POVMs: the tetrahedron ($m=4$), the octahedron ($m=6$), the hexahedron ($m=8$), the icosahedron ($m=12$), the dodecahedron ($m=20$), and the icosidodecahedron\footnote{A quasi-regular polyhedron} ($m=30$). The probabilities generated by the protocol are compared with the target Born probabilities and with the probabilities obtained by direct finite sampling from the Born rule.

\begin{figure}[htbp]
    \centering
    \includegraphics[width=8cm]{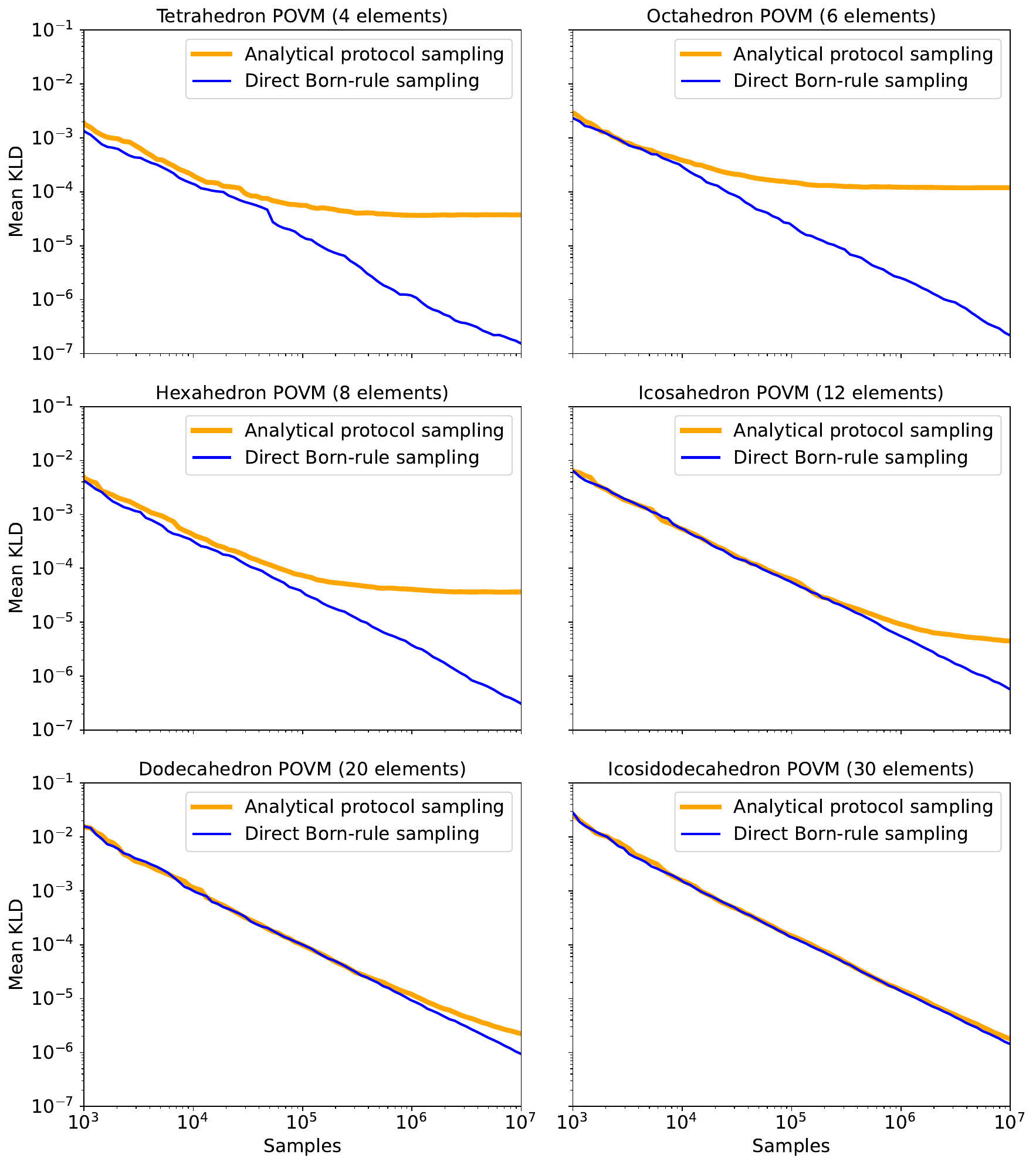}
    \caption{KLD of the analytical 1-bit protocol and of direct Born-rule sampling for six regular eIC-POVMs: tetrahedron, octahedron, hexahedron, icosahedron, dodecahedron, and icosidodecahedron ($m=4,6,8,12,20,30$, respectively). Each curve is averaged over $M=50$ randomly sampled input states. The blue curves represent the Born-rule sampling baseline, while the yellow curves represent the analytical 1-bit protocol.}
    \label{polyhedrons}
\end{figure}

In this section, we quantify the discrepancy using the Kullback-Leibler divergence(KLD), averaged over randomly chosen input states.
In contrast with the MAE used in Sec.~III to define the training loss and measures the
average absolute deviation between probabilities, the KLD is more sensitive to
the full shape of the probability distribution, and in particular to relative
errors in outcomes with small probabilities. For this reason, the numerical
values obtained from the KLD can differ significantly from those obtained with
the MAE. 
While both measures lead to neural networks that performed similarly both in training and testing, the extra sensitivity of the KLD is useful to us now to closely compare the complete output distributions generated by different protocols. The results are shown in Fig.~\ref{polyhedrons}.

For small sample sizes, the KLD is dominated by statistical fluctuations, and both curves decrease as the number of samples increases. The Born-rule sampling baseline continues to approach zero, whereas the KLD of the 1-bit protocol eventually reaches a nonzero plateau.
This plateau represents the intrinsic bias of the protocol with respect to the Born probabilities and shows that the finite-POVM simulation is not exact.

The agreement is nevertheless very accurate for all regular polyhedral configuration studied. Moreover the intrinsic discrepancy becomes small for the POVMs with larger and more isotropically distributed sets of Bloch vectors, such as the dodecahedral and the icosidodecahedral configurations.

\begin{figure}[htbp]
    \centering
    \includegraphics[width=\columnwidth]{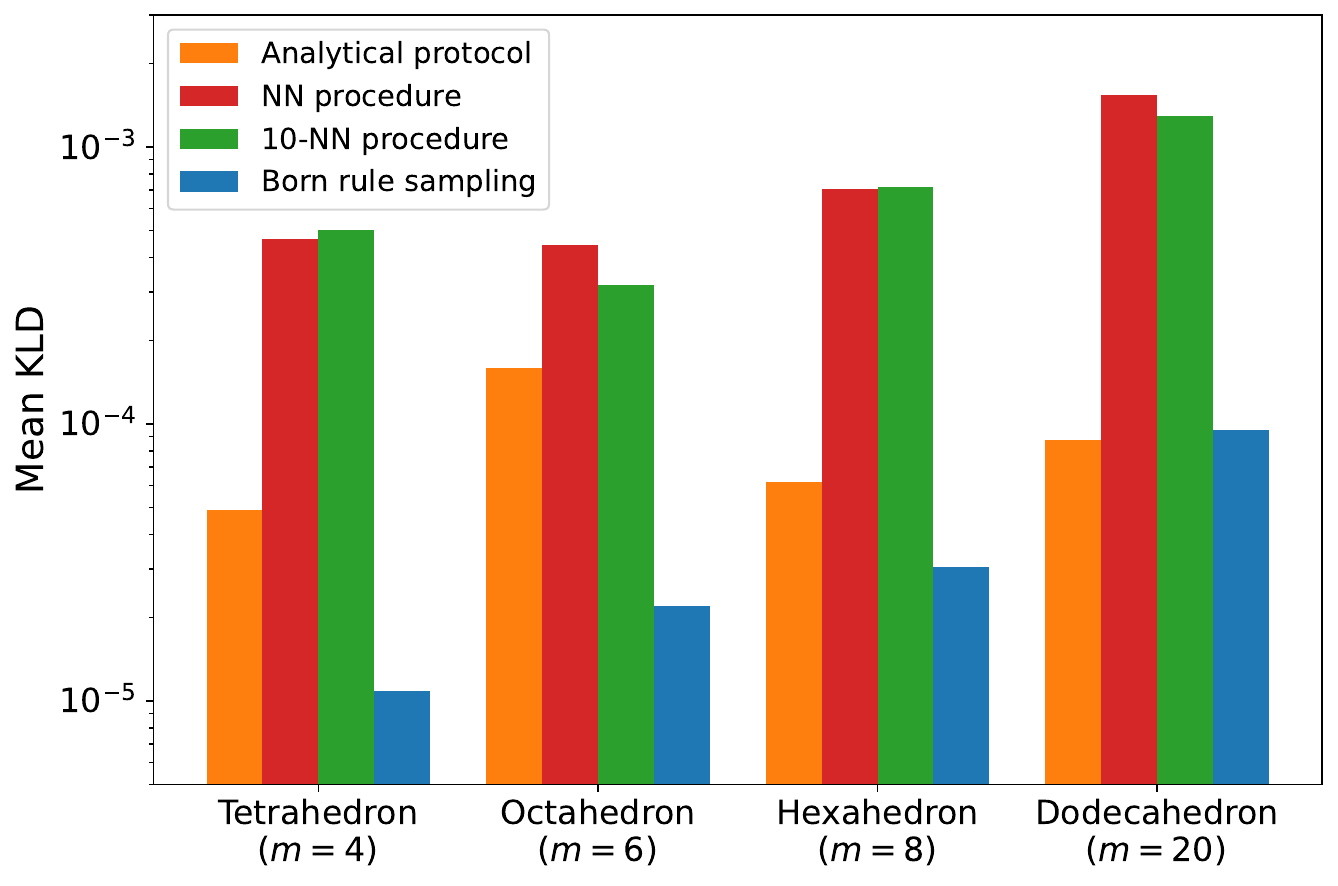}
    \caption{Mean KLD for the analytical protocol, the NN-procedure, an ensemble of ten NN-procedures, and direct sampling from the Born rule. For each configuration, the mean is computed over $M=20$ random input states $\vec r$, using $N=10^5$ samples per state.}
    \label{fig:comparison_protocols}
\end{figure}

We also compare the analytical 1-bit protocol directly with the neural network output.  Fig.~\ref{fig:comparison_protocols} shows the mean KLD of the analytical protocol of a single trained neural network, an ensemble average over $10$ independently trained neural networks, and direct sampling from the Born rule.

The analytical protocol gives a smaller KLD than both the single-network response and the ensemble-average neural network response, showing that the explicit rule in (\ref{eq:semi_analytical_protocol}) better reproduces the Born statistics.

These results indicate that, while the neural network learns the correct qualitative structure, the explicit analytical rule provides a more accurate realization of this structure at the level of the full output distribution. 
Note also that the averaged neural network response has a similar mean KLD than that of a single network. Despite not improving statistical accuracy, averaging over independently trained networks was key to distill the geometric response pattern condensed in Eqs.~\eqref{eq:nn_rule_c1} and \eqref{eq:nn_rule_c0}.

Fig.~\ref{fig:comparison_protocols} also shows that the KLD of the 1-bit protocol approaches the KLD obtained by direct sampling from the Born rule as the number of POVM elements increases. This trend is consistent with Fig.~\ref{polyhedrons}: at $10^5$ samples, the 1-bit protocol is 
already extremely close to the Born rule sampling benchmark, especially for the
dodecahedral and icosidodecahedral POVMs. This supports the interpretation that regular polyhedral eIC-POVMs with more elements provide better finite approximations to the isotropic continuum construction discussed below.

In conclusion, the 1-bit protocol improves on the numerical performance of the
trained NN-procedures while retaining the same qualitative feature: 1-bit
communication reproduces the target statistics with high accuracy for
eIC-POVMs.

\subsection{The continuous POVM limit}

The analytical expression for the probability distribution over outcomes of the 1-bit protocol takes the following form. 
The probability density of the shared vector $\vec{\lambda}$ after step 3 of the protocol [cf. (\ref{eq:sa_flip})] is 
\begin{equation}
    \rho (\vec{\lambda}| \vec{r})=2 H(\vec{r}\cdot\vec{\lambda}) \,,
\end{equation}
and the probability of obtaining outcome $i$ given a state with Bloch vector $\vec{r}$ reads [cf. (\ref{rtq})]
\begin{equation}\label{eq:final-prob}
P_{\mathrm{1-bit}}(i|\vec{r},\; Y)=2 \int d\vec{\lambda}
\frac{ H(\vec{r}\cdot\vec{\lambda}) \,(\vec{y}_i\cdot\vec{\lambda})\,H(\vec{y}_i\cdot\vec{\lambda})}
{\sum_{j=1}^{m}(\vec{y}_j\cdot\vec{\lambda})\,H(\vec{y}_j\cdot\vec{\lambda})},
\end{equation}
where  recall that $d\vec\lambda=d \Omega/4 \pi$ is the 
Haar measure in $\mathbb{S}^2$.
Although (\ref{eq:final-prob}) provides a remarkably accurate approximation to the eIC-POVM probabilities, it is not exact in general. This can be seen from a simple example. For instance, in the octahedral configuration, consider the probability of obtaining the outcome $+\hat z$ when Alice's state points along the diagonal in the $x$-$z$ plane. In this case, one can show analytically that the quantum probability and the probability generated by the 1-bit protocol have a discrepancy of $\mathcal{O}(10^{-3})$.
At the other extreme, if Alice's state happens to align with one of Bob's measurement outcomes, \emph{i.e.}, $\vec r=\vec{y}_i$ for some $i$, one can show that the 1-bit protocol is exact for all outcomes $i$ (See Appendix \ref{app: exactness}.).

This notwithstanding, in the limit of a continuous uniform POVM $\mathcal{M}=\{2\ketbrad{\vec m}d\vec m\}_{\vec m\in \mathbb{S}^2}$, $d\vec m = d\Omega/4\pi$, 
~(\ref{eq:final-prob})
yields the exact quantum probability density, which reads
\begin{equation}
\label{eq:exact_continuous}
   P_Q(\vec{m}| \rho_{\vec{r}}, \mathcal{M})= 1+\vec{r}\cdot \vec{m} \,.
\end{equation}
Indeed, by rotational symmetry the denominator in~(\ref{eq:final-prob}) becomes the number
\begin{equation}
\label{eq:denominator}
    \int d\vec{m} \, (\vec{m}\cdot\vec{\lambda})\,H(\vec{m}\cdot\vec{\lambda})=\frac{1}{4}.
\end{equation}

Using $x\,H(x)=(x+|x|)/2$, we get 
\begin{equation}
\label{eq:cont-prob}
P_{\mathrm{1-bit}}(\vec{m} |\vec{r}, Y)=4 \int d\vec{\lambda}
 H(\vec{r}\cdot\vec{\lambda})\left[ \,(\vec{m}\cdot\vec{\lambda}) + |\vec{m}\cdot\vec{\lambda}|\right] \,.
\end{equation}
 The first term is proportional to $\vec{r}\cdot \vec{m}$, while the second is again a number, \emph{i.e.}, $P_{\mathrm{1-bit}}(\vec{m} |\vec{r}, Y)=\gamma_1 + \gamma_2 \vec{r}\cdot \vec{m}$. Taking $\vec{m}=-\vec{r}$, the integral vanishes and one gets $\gamma_1= \gamma_2$. For  $\vec{m}= \vec{r}$ we get $\gamma_1=1$, and we recover (\ref{eq:exact_continuous}).

We have thus so far established that the 1-bit protocol provides accurate but not exact eIC-POVM statistics in general, and that it simulates exactly the statistics of the continuous POVM. 

We bridge these facts by giving evidence of the decreasing behavior of the $L_1$ distance $L_1(P_{\mathrm{1-bit}},P_Q)\coloneqq \sum_{j=1}^m |P_{\mathrm{1-bit}}(j|\vec{r},Y) - P_Q(j|\rho_{\vec{r}},\mathcal{M}_Y)|$ of the 1-bit protocol for eIC-POVMs with increasing number of outcomes. Note that MAE is the average $L_1$ distance over batch of inputs, hence the choice. We do this in two ways. For spherical $t$-designs (a subset of the class of eIC-POVMs, where $m\geq t^2/4$), we prove analytically that the $L_1$ distance scales as $\mathcal{O}(1/\sqrt{t})$ (see Appendix~\ref{app:scalingtdesigns}.). In addition, we perform a numerical worst-case search for the regular polyhedra considered in Fig.~\ref{polyhedrons}. That is, for each POVM configuration, we search for the state in the Bloch sphere that maximizes the $L_1$ distance when the PM scenario is simulated using the 1-bit protocol. Table~\ref{tab:mae_opt} provides the results.

\begin{table}[t] \centering

\begin{tabular}{| c|c || c  |}
 \hline
 \multicolumn{3}{|c|}{Worst-case $L_1(P_{\mathrm{1-bit}},P_Q)$ }\\
 \hline
 Configuration & $m$ & $L_1$   \\
 \hline
 Projective    & 2  & 0.2138  \\
 Tetrahedron   & 4  & 0.0093  \\
 Octahedron    & 6  & 0.0159 \\
 Hexahedron    & 8  & 0.0085  \\
 Icosahedron    & 12  & 0.0034  \\
 Dodecahedron  & 20 & 0.0029  \\
 Icosidodecahedron    & 30  & 0.0028  \\
 \hline
\end{tabular}
\caption{Worst-case $L_1$ distance for the measurement configurations considered. The order of magnitude of error, decreases by increasing the number of outcomes, starting from $\mathcal{O}(10^{-1})$ for $m=2$ up to $\mathcal{O}(10^{-3})$ for $m=30$.}
\label{tab:mae_opt}
\end{table}

\section{Conclusions}

We have shown that one bit of classical communication can approximate with high accuracy qubit prepare-and-measure statistics for restricted, highly structured families of measurements. Using a search by NN-procedure, we identified equal-trace rank-one POVMs---and, in particular, informationally complete symmetric configurations---as the measurement families most amenable to 1-bit classical simulation. The numerical results indicate that uniform weights and three-dimensional symmetry are the key features behind this enhanced simulability.

Guided by the structure learned by the network, we derived an explicit analytical 1-bit protocol and analyzed its performance on finite symmetric POVMs. Although the protocol is not exact for finite configurations, its error decreases as the measurement becomes more isotropic. In the continuous limit, corresponding to the covariant qubit POVM, the protocol becomes exact. This proves that the continuous isotropic qubit measurement is exactly 1-bit simulable in the prepare-and-measure scenario. 

Our results demonstrate that the 2-bit communication cost required for arbitrary qubit POVMs is not representative of all measurement families. Symmetry and isotropy can substantially reduce the classical resources needed to reproduce quantum statistics. This result loosely aligns with other recent observations pointing out that, perhaps contrary to prior common understanding, SIC-POVMs
do \emph{not} comprise the "most quantum'' representatives among all measurements:  they are not the hardest to simulate by convex combinations of projective measurements~\cite{cobucci2026}, and they possess the least intrinsic randomness~\cite{curran2026}. 

More broadly, this work contributes to the growing effort to identify low-communication classical descriptions of nonclassical correlations. Related questions have been studied in Bell-simulation settings, including 1-bit protocols, communication-assisted models, and neural network based approaches to nonclassicality problems \cite{montina2012epistemic,renner2023minimal,krivachy2020neural}. Our results show that neural networks can also help uncover analytically tractable 1-bit protocols in prepare-and-measure scenarios, and point toward a systematic classification of measurement families with reduced classical communication cost.

\section{Acknowledgements}
RMT, GS, and MZ acknowledge financial support from MCIN 
grant PID2022-141283NB-I00 funded by MCIN/AEI/10.13039/501100011033. MZ acknowledges Some Sankar Bhattacharya and support from the Ministerio de Ciencia e Innovación of the Spanish Government, contract PRE2020-093634. GG acknowledges DAAD, the Deutsche Forschungsgemeinschaft (DFG, German Research Foundation, project numbers 447948357 and 440958198), the Sino-German Center for Research Promotion (Project M-0294), and the German Ministry of Education and Research (Project QuKuK, BMBF Grant No. 16KIS1618K). GS acknowledges funding from the Programa Talent UAB - Banco de Santander. EF acknowledges support by the Generalitat de Catalunya and the European Social Fund grant Joan Oró 2025 FI-1 00848. JE acknowledges the funding received by i2CAT from Department de Recerca i Universitats of the Generalitat de Catalunya.

\bibliography{References}

\clearpage
\onecolumngrid

\appendix

\section{Neural network training methodology and hyperparameter selection}
\label{app:training}

\subsection{Training data and batch-averaged loss}

For a given POVM $\mathcal{M}_{Y}$, the training dataset is constructed as follows. First, we generate $M$ random pure qubit states sampled uniformly on the Bloch sphere. We denote their Bloch vectors by $\vec r^{(j)}$, with $j=1,\ldots,M$.

For each state $\vec r^{(j)}$, the target quantum probability distribution is computed using the Born rule,
\begin{equation}
P_Q(i|\rho_{\vec r^{(j)}},\mathcal{M}_Y)
=
p_i\bigl(1+\vec y_i\cdot\vec r^{(j)}\bigr),
\qquad
i=1,\ldots,m.
\end{equation}
These probabilities constitute the target distributions of the training procedure.

Next, for each state $\vec r^{(j)}$, we generate $N$ realizations of the shared randomness. We denote the $k$-th random vector associated with the state $\vec r^{(j)}$ by $\vec\lambda^{(j,k)}$, with $k=1,\ldots,N$. Each vector is sampled uniformly on the unit sphere. The corresponding communicated bit is then computed as
\begin{equation}
c^{(j,k)}
=
H\bigl(
\vec\lambda^{(j,k)}
\cdot
\vec r^{(j)}
\bigr).
\end{equation}

Therefore, each state $\vec r^{(j)}$ gives rise to $N$ neural network inputs, each of them formed by the three components of $\vec\lambda^{(j,k)}$ and the bit $c^{(j,k)}$. The full training dataset contains $M\times N$ input instances. However, these instances are naturally grouped into $M$ batches, each corresponding to one quantum state.

This grouping is essential because the neural network is not required to reproduce the quantum probability distribution for each individual realization of the shared randomness. Instead, the relevant quantity is the output distribution obtained after averaging over the $N$ realizations associated with the same state $\vec r^{(j)}$.

Let $\hat y_i^{(j,k)}$ denote the neural network prediction for outcome $i$ for the $k$-th shared randomness realization associated with the state $\vec r^{(j)}$. The batch-averaged prediction for that state is
\begin{equation}
\bar y_i^{(j)}
=
\frac{1}{N}
\sum_{k=1}^{N}
\hat y_i^{(j,k)} = \bar p_i^{(j)},
\end{equation}
where $\bar p_i^{(j)}$ is the same as (\ref{eq:nn_mc_average}). This quantity is compared with the target quantum probability
\begin{equation}
y_i^{(j)}
:=
P_Q(i| \rho_{\vec r^{(j)}},\mathcal{M}_Y).
\end{equation}
The loss associated with the state $\vec r^{(j)}$ is defined by 
\begin{equation}
L_j
=
\sum_{i=1}^{m}
\left[\bar p_{i}^{(j)}-P_Q(i|\rho_{\vec{r}^{(j)}},\mathcal{M}_Y)\right]^2.
\label{eq:Lj}
\end{equation}
The full loss function, according to (\ref{eq:nn_mse_loss}) is then obtained by averaging over the $M$ sampled states:
\begin{equation}
\mathcal L_{\rm{MSE}}
=
\frac{1}{M}
\sum_{j=1}^{M}
L_j
=
\frac{1}{M}\sum_{j=1}^{M}\sum_{i=1}^{m}
\left[\bar p_{i}^{(j)}-P_Q(i|\rho_{\vec{r}^{(j)}},\mathcal{M}_Y)\right]^2.
\label{eq:loss}
\end{equation}

In this way, $M$ is the number of quantum states used during training, whereas $N$ is the number of shared randomness samples used for each state. Increasing $M$ improves the coverage of the Bloch sphere during training, while increasing $N$ improves the numerical estimate of the averaged output distribution.

\subsection{Network architecture and optimization}

For each POVM, we train an independent feed-forward neural network. The input layer has four neurons, corresponding to the three Cartesian components of $\vec\lambda$ and the communicated bit $c$. The network has two hidden layers with 10 and 16 neurons, respectively. The output layer has $m$ neurons, where $m$ is the number of outcomes of the POVM.

ReLU activation functions are used in the hidden layers. The output layer uses a softmax activation function, ensuring that the neural network output is a normalized probability distribution over the $m$ possible outcomes.

\begin{figure}[htbp]
\centering
\includegraphics[width=0.5\textwidth]{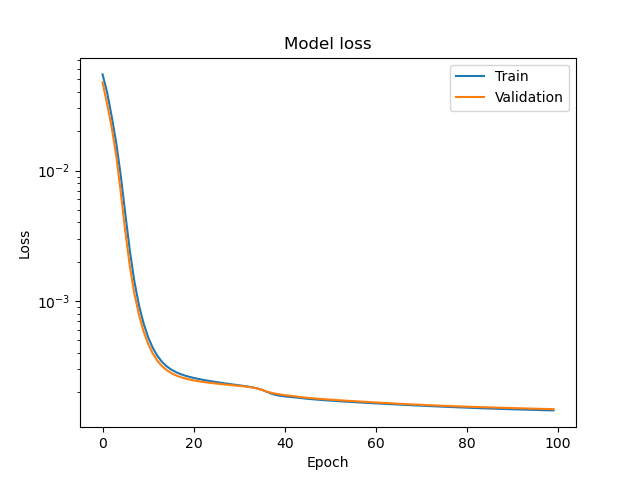}
\caption{Loss as a function of the number of epochs for the training and validation datasets.}
\label{Loss}
\end{figure}

The training-set size is determined by $M$ and $N$, corresponding to the number of randomly produced states, and number of independent Monte Carlo samples of shared randomness used per each state, respectively.  For the number of input states, we set $M_{\mathrm{val}}=\frac{M}{4}$ for validation, and $M_{\mathrm{test}}=1000$ for testing. For $N$, in both validation and testing phases, we have $N_{\mathrm{test}} = N_{\mathrm{val}}=N$. The network is trained using the Adam optimizer with a learning rate of $0.001$. In all simulations, the number of training epochs is fixed to 100.
An example of the training convergence is shown in Fig.~\ref{Loss}. The loss defined in (\ref{eq:loss}) converges to approximately $1.5\times10^{-3}$ after 100 epochs for both the training and validation datasets.

\subsection{Performance metric and selection of \texorpdfstring{$\bm{M}$}{M} and \texorpdfstring{$\bm{N}$}{N}}

The performance of the NN-procedure is evaluated using MAE between the batch-averaged neural network prediction and the target quantum probability distribution:
\begin{equation}
\mathrm{MAE}
=
\frac{1}{M}
\sum_{j=1}^{M}
\sum_{i=1}^{m}
\left|
\frac{1}{N}
\sum_{k=1}^{N}
\hat y_i^{(j,k)}
-
y_i^{(j)}
\right| = \frac{1}{M}
\sum_{j=1}^{M}
\sum_{i=1}^{m} \left|\bar p_{i}^{(j)}-P_Q(i|\rho_{\vec{r}^{(j)}},\mathcal{M}_Y)\right|.
\label{eq:mae}
\end{equation}
As in the loss function, the average over the $N$ shared randomness realizations is necessary because the relevant quantity is the probability distribution predicted for each state, rather than the output of the network for a single realization of $\vec\lambda$.

\begin{figure}[htbp]
\centering
\includegraphics[width=0.5\textwidth]{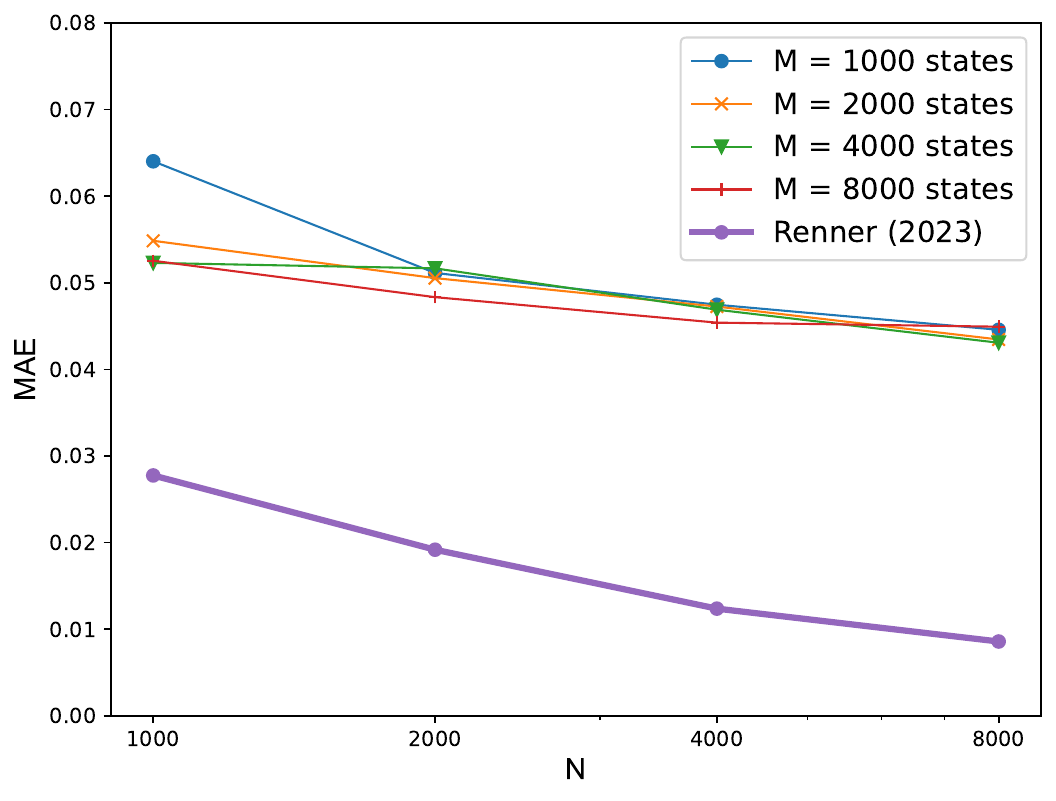}
\caption{MAE as a function of the batch size $N$ for different sample sizes $M$, compared with the MAE of the protocol of Renner \emph{et al.}~\cite{renner2023classical}.}
\label{MAE M N}
\end{figure}

To select suitable values of $M$ and $N$, we analyze the performance of the NN-procedure for a POVM with $m=6$ elements. Fig.~\ref{MAE M N} shows the MAE for different values of $M$ and $N$. The error decreases on average as both parameters increase, although the improvement is more pronounced when increasing $N$. This is expected because a larger value of $N$ gives a more accurate estimate of the batch-averaged distribution that the network is trained to reproduce, whereas a larger value of $M$ improves the coverage of the state space during training.

For comparison, Fig.~\ref{MAE M N} also includes the MAE obtained from the exact 2-bit protocol of Ref.~\cite{renner2023classical}. Although that protocol reproduces the quantum probabilities exactly in the infinite-sampling limit, its numerical implementation also depends on the number of samples used to estimate the corresponding probabilities. Therefore, finite-$N$ effects are also present in that case.

The comparison shows that the 1-bit NN-procedure remains less accurate than the exact 2-bit simulation, but nevertheless provides a nontrivial approximation to the target quantum statistics.

For all remaining experiments, we fix $N=4000$, and $M=2000$, which provides a good compromise between numerical accuracy and computational cost.

\section{Generation of random qubit POVMs}

Random rank-one qubit POVMs are generated by sets of random points, $\vec{y}$ in the unit sphere $\mathbb{S}^2$. These can be simply obtained from sampling two independent random variables $u,v \sim \mathcal{U}(0,1)$ and defining $\phi = 2\pi u$ and $\theta = \arccos(1-2v)$. The random unit vector in spherical coordinates $\vec{y} =
(\sin\theta \cos\phi,\,
 \sin\theta \sin\phi,\,
 \cos\theta)$ is then assured to be uniformly sampled.

In the implementation, we generated a  pool of $10^6$ points. For a POVM with $m$ outcomes, we then selected $m$ directions $\{\vec{y}_j\}_{j=1}^m$ at random from this pool. 
Once the directions $\{\vec{y}_j\}_{j=1}^m$ were fixed, the weights $\{p_j\}_{j=1}^m$ were obtained numerically by enforcing the POVM conditions: 
\begin{equation}
\sum_{j=1}^m p_j = 1,
\qquad
\sum_{j=1}^m p_j  \vec{y}_j = \vec{0}.
\end{equation}
We used the numerical routine L-BFGS-B~\cite{zhu1997algorithm} that allowed us to fix the initial point to $p_j=1/m$ and to enforce that no weight became too small. If a sampled set of directions did not admit an admissible solution, a new set of directions was generated and the procedure was repeated.
\section{Exactness of the protocol for the regular polyhedra, where the state aligns with one the measurement outcomes.}\label{app: exactness}
In this appendix, we prove that for the regular polyhedra, the protocol gives exact solution whenever $\vec{r}=\vec{y}_i$ for some $i$. 

We start from the analytical description of the protocol from (\ref{eq:final-prob}) of the main text:
\begin{align}
    P_{\mathrm{1-bit}}(k|\vec{r}=\vec{y}_i,Y) =2\int_{\mathbb{S}^2} d\vec{\lambda} \frac{H(\vec{y}_i\cdot\vec{\lambda})\vec{y}_k \cdot \vec{\lambda}\:H(\vec{y}_k\cdot\vec{\lambda})}{\sum_{j=1}^m \vec{y}_j \cdot \vec{\lambda}\:H(\vec{y}_j\cdot\vec{\lambda})}.
\end{align}
Now, by $\sum_{j=1}^m \vec{y}_j = 0$, using the identity $xH(x) = \frac{|x|+x}{2}$, and applying $H(\vec{y}_i\cdot\vec{\lambda})$, over the integral region, we get:
\begin{align}\label{eq: I_1 + I_2}
     P_{\mathrm{1-bit}}(k|\vec{r}=\vec{y}_i,Y) = 2\int_{\vec{y}_i \cdot \vec
     \lambda \geq 0} d\vec{\lambda} \: \frac{ |\vec{y}_k \cdot \vec{\lambda}|}{\sum_{j=1}^m |\vec{y}_j \cdot \vec{\lambda}|} + 2\int_{\vec{y}_i \cdot \vec
     \lambda \geq 0} d\vec{\lambda} \: \frac{\vec{y}_k \cdot \vec{\lambda}}{\sum_{j=1}^m |\vec{y}_j \cdot \vec{\lambda}|}\coloneqq  I_1 +  I_2.
\end{align}
Because the Integrand in $I_1$ is even under $\vec{\lambda}\mapsto -\vec{\lambda}$, we have:
\begin{align}\label{eq: I_1 int}
    I_1 =   \int_{\mathbb{S}^2} d\vec{\lambda} \frac{ |\vec{y}_k \cdot \vec{\lambda}|}{\sum_{j=1}^m |\vec{y}_j\cdot \vec{\lambda}|}= \frac{1}{m}
\end{align}
where we have used the symmetry of the regular polyhedra. For $I_2$ we have:
\begin{align}\label{eq: I_2 int}
    I_2 = 2 \vec{y}_k \cdot \:\int_{\vec{y}_i \cdot \vec{
    \lambda} \geq 0} d\vec{\lambda} \frac{\vec{\lambda}}{\sum_{j=1}^m |\vec{y}_j \cdot \vec{\lambda} |} \coloneqq 2 \vec{y}_k \cdot \vec{w},
\end{align}
where $\vec{w}$ is an effective vector. Now, let's see how it behaves under the symmetry rotation $R_i\vec{w}$, where $R_i$ is the nontrivial rotation around the axis $\vec{y}_i=R_i\vec{y}_i$, that maps the regular polyhedral shape to itself, up to relabeling the indices:
\begin{align}
    R_i \vec{w} = \int_{\vec{y}_i \cdot \vec{\lambda} \geq 0} d \vec{\lambda} \frac{R_i \vec{\lambda}}{\sum_{j=1}^m |\vec{y}_j \cdot \vec{\lambda}|} = \int_{(\vec{y}_i^T R_i^{T} R_i \vec{\lambda})\geq 0} |det{(R_i)}| d\vec{\lambda} \frac{R_i \vec{\lambda}}{\sum_{j=1}^m |\vec{y}_j^T R_i^{T} R_i \vec{\lambda}|} =\vec{w}.
\end{align}
The last equality results from changing the variable of integration by $\vec{\lambda}^\prime \coloneqq R_i \vec{\lambda}$ and using the fact $|det(R_i)|=1$. Therefore, we see that $\vec{w}= \alpha \vec{y}_i$. To derive $\alpha$, we apply the dot product with $\vec{y}_i$:
\begin{align}
    \vec{y}_i \cdot \vec{w} = \alpha = \int_{\vec{y}_i \cdot \vec{\lambda}\geq 0} d\vec{\lambda} \frac{\vec{y}_i \cdot \vec{\lambda}}{\sum_{j=1}^m |\vec{y}_j \cdot \vec{\lambda}|} = \frac{1}{2m},
\end{align}
where we have used the fact that in the region of integration, $\vec{y}_i \cdot \vec{\lambda} = |\vec{y}_i \cdot \vec{\lambda}|$, and (\ref{eq: I_1 int}). Now, by putting it into (\ref{eq: I_2 int}), we have:
\begin{align}
    I_2 = \frac{1}{m}\vec{y}_k \cdot \vec{y}_i,
\end{align}
which alongside with $I_1$ can be combined and substituted into (\ref{eq: I_1 + I_2}), which leads to the expression:
\begin{align}
    P_{\mathrm{1-bit}}(k|\vec{r}=\vec{y}_i,Y) = \frac{1}{m}(1 + \vec{y}_i \cdot \vec{y}_k) = P_Q(k|\rho_{\vec{r}=\vec{y}_i},\mathcal{M}_Y). \quad  \quad \quad \quad \square
\end{align}

\section{Scaling of \texorpdfstring{$\bm{L_1}$}{L1} distance for spherical \texorpdfstring{$\bm{t}$}{t}-designs}\label{app:scalingtdesigns}
In this appendix, we study the asymptotic behavior of $L_1$ distance of the analytical 1-bit protocol from quantum predictions, and provide an upper bound for spherical $t-$design measurements that scales as $\mathcal{O}(1/\sqrt{t})$.

As any spherical $t-$design is a particular example of an e-POVM, we start by writing:
\begin{align}
M_{i}
=
\frac{1}{m}
\left(
I+\vec{y}_{i}\cdot\vec{\sigma}
\right),
\qquad
\vec{y}_{i}\in \mathbb{S}^{2},
\end{align}
with $\sum_{i=1}^{m}\vec{y}_{i}=\vec{0}$. Without loss of generality, we also restrict to pure input states, so that $\vec{r}\in \mathbb{S}^{2}$. The Born rule probability is then
\begin{align}
P_{Q}(i|\rho_{\vec{r}},\mathcal{M}_Y)
=
\frac{1}{m}
\left(
1+\vec{y}_{i}\cdot\vec{r}
\right).
\end{align}
It is useful to write the quantum probabilities in integral form as
\begin{align}
\frac{1}{m}
\left(
1+\vec{y}_{i}\cdot\vec{r}
\right)=\frac{8}{m}\int d\vec{\lambda'}\, H(\vec{r} \cdot \vec{\lambda}') \vec{y}_i\cdot \vec{\lambda}'\, H(\vec{y}_i\cdot \vec{\lambda}')
= \frac{8}{m}\int_{\vec{r}\cdot\vec{\lambda}' \geq 0} d\vec{\lambda'}\,\vec{y}_i\cdot \vec{\lambda}'\, H(\vec{y}_i\cdot \vec{\lambda}')\,  .
\end{align}

Recall that the 1-bit protocol probability (\ref{eq:final-prob}) reads
\begin{equation}
P_{\mathrm{1-bit}}(i|\vec{r},\; Y)=2 \int d\vec{\lambda}'
\frac{ H(\vec{r}\cdot\vec{\lambda}') \,(\vec{y}_i\cdot\vec{\lambda}')\,H(\vec{y}_i\cdot\vec{\lambda}')}
{\sum_{j=1}^{m}(\vec{y}_j\cdot\vec{\lambda}')\,H(\vec{y}_j\cdot\vec{\lambda}')},
\end{equation}
where the denominator can be written as
\begin{align}
\label{eq: S_Y definition}
\sum_{j=1}^{m}
(\vec{y}_{i}\cdot\vec{\lambda}^\prime) H(\vec{y}_{i}\cdot\vec{\lambda}^\prime)=
\frac{1}{2} \sum_{j=1}^{m}|\vec{y}_{j}\cdot\vec{\lambda}^\prime|
\coloneq S_{Y}(\vec{\lambda}')
\end{align}
In the above we have used the identity  $xH(x)=(x+|x|)/2$ and the $t$-design property $\sum_i\vec{y_i}=0$.

It is also convenient to define the quantities 
\begin{align}
Q(i|\vec{\lambda}^\prime,Y)
=
\frac{4}{m}
(\vec{y}_{i}\cdot\vec{\lambda}^\prime) H(\vec{y}_{i}\cdot\vec{\lambda}^\prime),
\end{align}
and 
\begin{align}
Z_{Y}(\vec{\lambda}^\prime)
:=
\sum_{i=1}^{m}
Q(i|\vec{\lambda}^\prime,Y)
=
\frac{2}{m} \sum_{j=1}^{m}|\vec{y}_{j}\cdot\vec{\lambda}^\prime|
=
\frac{4}{m}
S_{Y}(\vec{\lambda}^\prime),
\end{align}
to have 
\begin{align}
|P_{\mathrm{1-bit}}(i|\vec{r},Y)
-
P_{Q}(i|\rho_{\vec{r}},\mathcal{M}_Y)|
=
2\left|
\int_{\vec{r}\cdot\vec{\lambda}^\prime\geq 0}
Q(i|\vec{\lambda}^\prime,Y)
\left[
\frac{1}{Z_{Y}(\vec{\lambda}^\prime)}
-
1
\right]
d \vec{\lambda}^\prime.\right|
\leq
2
\int_{\vec{r}\cdot\vec{\lambda}^\prime\geq 0}
Q(i|\vec{\lambda}^\prime,Y)
\left|
\frac{1}{Z_{Y}(\vec{\lambda}^\prime)}
-
1
\right|
d \vec{\lambda}^\prime.
\label{eq:abs_bound}
\end{align}

We now use the expansion of the absolute function $|x|$ in terms of a series of Legendre polynomials 
\begin{align}
    |x| = \sum_{k=0}^\infty c_{2k} P_{2k}(x),
\end{align}
where the coefficients read

\begin{align}\label{eq:coefficients}
    c_{2k} =(-1)^{k-1} (4k+1)  \frac{(2k-3)!!}{(2k+2)!!},
\end{align}
(note that by analytical extension $c_0=1/2$) to write the terms $|\vec{y}_i \cdot \vec{\lambda}^\prime|$ in (\ref{eq: S_Y definition}) in the Legendre polynomials basis:
\begin{align}
    S_{Y}(\vec{\lambda}^\prime) =\frac{1}{2}\sum_{j=1}^m [\sum_{k=0}^\infty c_{2k}P_{2k}(\vec{y}_j \cdot \vec{\lambda}^\prime)].
\end{align}\label{eq:legendre_expansion}
Notice that for spherical $t-$designs one has:
\begin{align}
    &S_{Y}(\vec{\lambda}^\prime)=\frac{1}{2}\sum_{j=1}^m \Big( \sum_{k=0}^{\lfloor t/2 \rfloor} c_{2k}P_{2k}(\vec{y}_j \cdot \vec{\lambda}^\prime) +\sum_{k=\lfloor t/2 \rfloor+1}^\infty c_{2k}P_{2k}(\vec{y}_j \cdot \vec{\lambda}^\prime)\Big) \notag \\
    &= \frac{m}{2} \int_{\mathbb{S}^2} d \vec{y}  \:\Big(\sum_{k=0}^{\lfloor t/2 \rfloor}c_{2k}P_{2k}(\vec{y} \cdot \vec{\lambda}^\prime)\Big) + \frac{1}{2}\sum_{j=1}^m\sum_{k=\lfloor t/2 \rfloor+1}^\infty c_{2k}P_{2k}(\vec{y}_j \cdot \vec{\lambda}^\prime)\notag \\
    &=\frac{m}{2} \;\mathbb{E}_{\vec{y}}\Big[\Big(\sum_{k=0}^{\lfloor t/2 \rfloor}c_{2k}P_{2k}(\vec{y} \cdot \vec{\lambda}^\prime)\Big)\Big] +\frac{1}{2}\sum_{j=1}^m\sum_{k=\lfloor t/2 \rfloor+1}^\infty c_{2k}P_{2k}(\vec{y}_j \cdot \vec{\lambda}^\prime).
\end{align}
We note that all Legendre polynomials up to $\lfloor{t/2}\rfloor$ satisfy $\mathbb{E}[P_{2k}(\vec{y}\cdot \vec{\lambda}^\prime)]=0$, except for $k=0$, which gives $\mathbb{E}[P_0(\vec{y}\cdot \vec{\lambda}^\prime)]=1$, resulting in the first term reproducing a constant value of $m/4$, 
and therefore, one has:
\begin{align}
    S_{Y}(\vec{\lambda}^\prime) = \frac{m}{4} + R_t(\vec{\lambda}^\prime),
\end{align}
where
\begin{align}
    R_t(\vec{\lambda}^\prime):=\frac{1}{2}\sum_{j=1}^m \sum_{k=\lfloor t/2 \rfloor + 1}^\infty c_{2k} P_{2k}(\vec{y}_j \cdot \vec{\lambda}^\prime).
\end{align}
Additionally, as:
\begin{align}\label{eq:R_upper_bound}
    \frac{4}{m}|R_t(\vec{\lambda}^\prime)| =\frac{2}{m}\Bigg|\sum_{j=1}^m |\vec{y}_j\cdot \vec{\lambda}^\prime| - \frac{m}{2} \Bigg| < 1,
\end{align} 
substituting it into (\ref{eq:abs_bound}) leads to:
\begin{align}\label{eq:upper_bound_individual_P}
   \left|
P_{\mathrm{1-bit}}(i|\vec{r},Y)
-
P_{Q}(i|\rho_{\vec{r}},\mathcal{M}_Y)
\right|
&\leq 
2
\int_{\vec{r}\cdot\vec{\lambda}^\prime\geq 0}
Q(i|\vec{\lambda}^\prime,Y)
\left|
\frac{1}{(1+\frac{4}{m} R_t(\vec{\lambda}^\prime))}
-
1
\right| d \vec{\lambda}^\prime \notag \\
&\leq \mathrm{sup}_{\vec{\lambda}^\prime}\left|  \frac{4R_t(\vec{\lambda}^\prime)}{m+4R_t(\vec{\lambda}^\prime)}\right| P_Q(i|\rho_{\vec{r}},\mathcal{M}_Y).
\end{align}
Now, let's analyze $F(R_t(\vec{\lambda}^\prime)):= |\frac{4R_t(\vec{\lambda}^\prime)}{m + 4R_t(\vec{\lambda}^\prime)}|$, as a function of $R_t$. To find an upper bound, from (\ref{eq:R_upper_bound}) it follows that: 
\begin{align}\label{eq:sup}
    \textrm{sup}_{\vec{\lambda}^\prime\in \mathbb{S}^2}[F(R_t(\vec{\lambda}^\prime)] \leq \frac{ 4 \:\textrm{sup}_{\vec{\lambda}^\prime\in \mathbb{S}^2}[|R_t(\vec{\lambda}^\prime)|]}{ m - 4 \:\textrm{sup}_{\vec{\lambda}^\prime\in \mathbb{S}^2}[|R_t(\vec{\lambda}^\prime)|]   }
\end{align}
On the other hand: 
\begin{align}
   |R_t(\vec{\lambda}^\prime)| &\leq \frac{1}{2}\sum_{j=1}^m \sum_{k=\lfloor t/2 \rfloor+1}^\infty |c_{2k}| \cdot |P_{2k}(\vec{y}_j \cdot \vec{\lambda}^\prime)|\leq \frac{m}{2} \sum_{k=\lfloor t/2 \rfloor+1}^\infty |c_{2k}|
\end{align}
as $|P_{2k}(\cdot)| \leq 1$ for all $k$'s.
Substituting this inequality into (\ref{eq:sup}), for large enough $t$, where $\sum_{k=\lfloor t/2 \rfloor+1}^\infty |c_{2k}| <1/2$, results in:
\begin{align}\label{eq: bound_sup}
    \textrm{sup}_{\vec{\lambda}^\prime\in \mathbb{S}^2}[F(R_t(\vec{\lambda}^\prime)] \leq \frac{2\:\textrm{sup}_{\vec{\lambda}^\prime\in \mathbb{S}^2}|R_t(\vec{\lambda}^\prime)|}{ 1 - 2\:\textrm{sup}_{\vec{\lambda}^\prime\in \mathbb{S}^2}|R_t(\vec{\lambda}^\prime)| } \leq \frac{2\sum_{k=\lfloor t/2 \rfloor+1}^\infty |c_{2k}|}{1 - 2\sum_{k=\lfloor t/2 \rfloor+1}^\infty |c_{2k}|. }
\end{align}\label{eq:bound_sup}
We finally upper bound the coefficients $|c_{2k}|$. For this, we rewrite (\ref{eq:coefficients}) as:
\begin{align}
    |c_{2k}|=\frac{4k +1}{4k^2 + 2k -2}\cdot \frac{(2k-1)!!}{(2k)!!}
\end{align}
and treat the first and second term separately. For the first term, we have: 
\begin{align}\label{eq: c_abs terms}
    \frac{4k+1}{4k^2 + 2k -2}= \frac{5}{4} \quad \textrm{for $k=1$, \ and} \
    \ \ \frac{4k+1}{4k^2 + 2k -2} \leq \frac{1}{k}  \quad \textrm{for $k>1$}.
\end{align}
For the second term,  we use the Wallis identity
\begin{align}
\label{eq:wallis}
    \frac{2}{\pi}\int_0^{\pi/2} dx \cos^{2k} x=\frac{(2k-1)!!}{(2k)!!}\, ,
\end{align}
together with the inequality $\log(\cos x)\leq -x^2/2$, to obtain
\begin{align}  
\label{eq:log-trick}
\frac{(2k-1)!!}{(2k)!!}\leq\frac{2}{\pi}\int_0^{\pi/2} dx \, e^{-k x^2}< \int_0^{\infty} dx \, e^{-k x^2}=\frac{1}{\sqrt{\pi k}}
\end{align}

Consequently, each term of (\ref{eq: c_abs terms}) is upper bounded by:
\begin{align}
    |c_2| \leq \frac{5}{4\sqrt{\pi}}; \quad |c_{2k >2}| \leq \frac{1}{\sqrt{k^{3} \pi}}.
\end{align}
Further, 
the summation $\sum_{k=\lfloor t/2\rfloor +1}^\infty |c_{2k}|$ is further upper bounded by using the fact that $\frac{1}{\sqrt{k^3\pi}}$ is a monotonic decreasing function of $k$ 
and the relationship between the sum and the corresponding integral 
\begin{align}
    \sum_{k=\lfloor t/2 \rfloor + 1}^\infty |c_{2k}| \leq \frac{1}{\sqrt{\pi}} \int_{\lfloor t/2 \rfloor}^\infty k^{-3/2} dk = \frac{2}{\sqrt{\pi}\sqrt{\lfloor t/2 \rfloor}}.
\end{align}
Therefore, by putting it into (\ref{eq: bound_sup}), one gets:
\begin{align}
        \textrm{sup}_{\vec{\lambda}^\prime\in \mathbb{S}^2}[F(R_t(\vec{\lambda}^\prime)] \leq \frac{4}{ \sqrt{\pi}\sqrt{\lfloor t/2 \rfloor}}\cdot\frac{ \sqrt{\pi}\sqrt{\lfloor t/2 \rfloor}}{ \sqrt{\pi}\sqrt{\lfloor t/2 \rfloor} - 4} = \frac{4}{ \sqrt{\pi}\sqrt{\lfloor t/2 \rfloor} - 4}.
\end{align}
and further, by substituting it into (\ref{eq:upper_bound_individual_P}), one has: 
\begin{align}
|
P_{\mathrm{1-bit}}(i|\vec{r},Y)
-
P_{Q}(i|\rho_{\vec{r}},\mathcal{M}_Y)
|
&\leq 
\frac{4}{ \sqrt{\pi}\sqrt{\lfloor t/2 \rfloor} - 4} P_Q(i|\rho_{\vec{r}},\mathcal{M}_Y).\label{eq:error_bound_per_outcome}
\end{align}
Finally, by summing over all outcomes, 
the upper bound for $L_1(P_{\mathrm{1-bit}},P_Q)$ reads:

\begin{align}
L_1(P_{\mathrm{1-bit}},P_Q) = \sum_{i=1}^m  \left|
P_{\mathrm{1-bit}}(i|\vec{r},Y)
-
P_{Q}(i|\rho_{\vec{r}},\mathcal{M}_Y) 
\right| &\leq \frac{4}{ \sqrt{\pi}\sqrt{\lfloor t/2 \rfloor} - 4} = \mathcal{O}(1/\sqrt{t}) \quad \textrm{as }t\rightarrow \infty.  \quad\square
\end{align}

\section{Additional Plots}
\label{app:additional_plots}

The main text focuses on a representative conditional response map, namely the
outcome $\vec{y}_5$ for the transmitted bit $c=1$, in order to illustrate how the
ensemble-averaged neural network output motivates the analytical 1-bit protocol.
Here, we provide the corresponding complete set of response maps for the
six-direction Cartesian (octahedral) eIC-POVM. In particular, we display all
six outcomes and both possible values of the transmitted bit.

Each figure is organized in six rows and four columns. Row $i$ corresponds to
the POVM element $\vec{y}_i$, while the first three columns show the
neural network conditional response probabilities obtained from ensembles of
size $k=1$, $k=5$, and $k=100$, respectively. The last column displays the
corresponding response of the analytical 1-bit protocol defined in (\ref{eq:semi_analytical_protocol}). In each panel, the probability is
shown as a function of the shared random vector $\vec{\lambda}$, parametrized
by the polar angles $(\phi,\theta)$. The red marker indicates the direction of
the Bloch vector associated with the outcome shown in that row. The common
color scale, ranging from $0$ to $1$, allows a direct comparison of the
different ensemble sizes, outcomes, and analytical predictions.

\begin{figure*}[t]
\centering
\includegraphics[
width=\textwidth,
height=0.82\textheight,
keepaspectratio
]{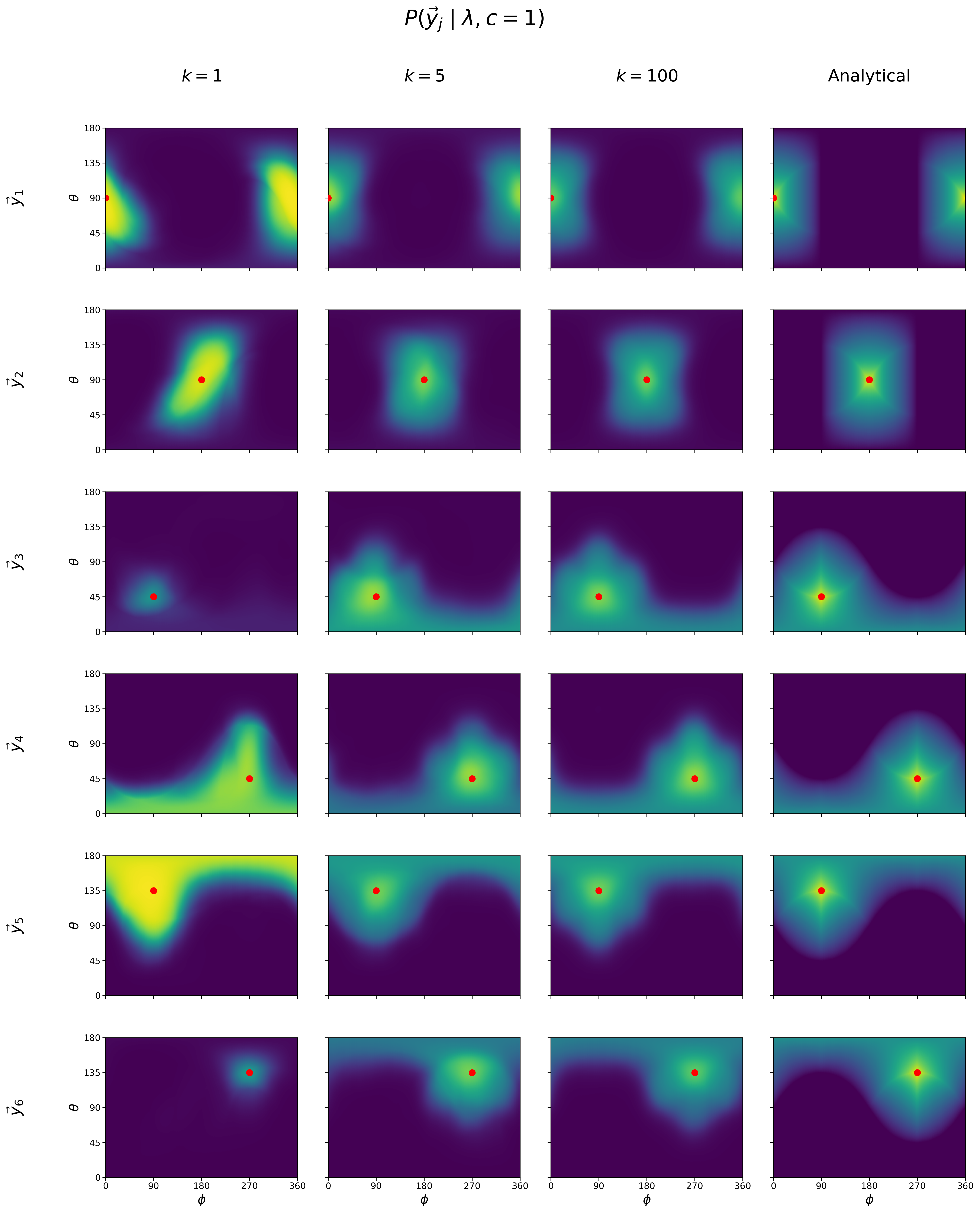}
\caption{
Conditional response maps for the six outcomes of the six-direction
Cartesian (octahedral) eIC-POVM, conditioned on the transmitted bit
$c=1$. Each row corresponds to one POVM element $\vec{y}_i$, with
$i=1,\ldots,6$. The first three columns show the neural network response
probabilities $P_{\mathrm{NN}}(i|c=1,\vec{\lambda},Y)$ for ensemble
sizes $k=1$, $k=5$, and $k=100$, respectively. The fourth column shows
the analytical response
$P_{\mathrm{1-bit}}(i|\vec{\lambda}^{,\prime},Y)$ from
(\ref{eq:semi_analytical_protocol}), where $c=1$. The red marker in
each panel identifies the Bloch vector $\vec{y}_i$ associated with the
displayed outcome.}
\label{fig:additional_maps_c1}
\end{figure*}

\begin{figure*}[t]
\centering
\includegraphics[
width=\textwidth,
height=0.82\textheight,
keepaspectratio
]{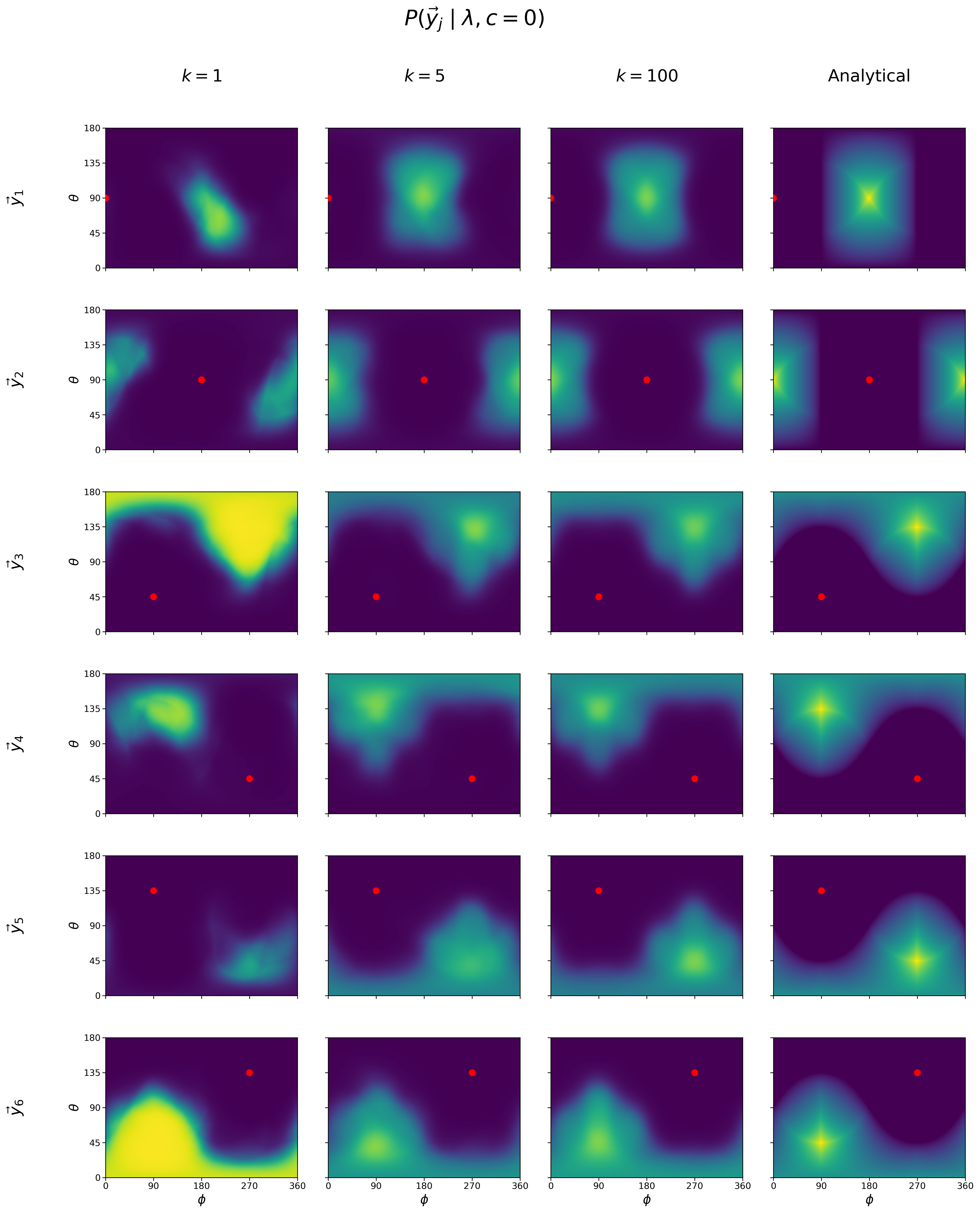}
\caption{
Conditional response maps for the six outcomes of the six-direction
Cartesian (octahedral) eIC-POVM, conditioned on the transmitted bit
$c=0$. The arrangement is identical to that in
Fig.~\ref{fig:additional_maps_c1}: rows correspond to the outcomes
$\vec{y}_1,\ldots,\vec{y}_6$, and columns show the neural network
responses for ensemble sizes $k=1$, $k=5$, and $k=100$, followed by the
analytical 1-bit response. In this case, the analytical protocol uses the
flipped hidden variable
$\vec{\lambda}^{\prime}=-\vec{\lambda}$, as prescribed by
(\ref{eq:sa_flip}). The maps show that the same response structure is
recovered for $c=0$, with the support of each outcome shifted according
to this sign flip. Red markers indicate the directions of the
corresponding POVM Bloch vectors.}
\label{fig:additional_maps_c0}
\end{figure*}

\end{document}